\documentclass[reprint]{JASA}

\usepackage{algorithm2e}
\usepackage{algorithmicx}
\usepackage{algcompatible}

\begin{document}

\title[]{Solving Room Impulse Response Inverse Problems Using Flow Matching with Analytic Wiener Denoiser}


\author{Kyung Yun Lee}
\author{Nils Meyer-Kahlen}
\affiliation{Acoustics Lab, Dpt. of Information and Communications Engineering, Aalto University, Espoo, Finland}
\author{Sebastian J. Schlecht}
\affiliation{Multimedia Communications ~\& Signal Processing, Friedrich-Alexander-Universität Erlangen-Nürnberg (FAU), Erlangen, Germany}
\author{Vesa Välimäki}
\affiliation{Acoustics Lab, Dpt. of Information and Communications Engineering, Aalto University, Espoo, Finland}




\email{kyung.y.lee@aalto.fi}




\begin{abstract}
Room impulse response (RIR) estimation naturally arises as a class of inverse problems, including denoising and deconvolution. While recent approaches often rely on supervised learning or learned generative priors, such methods require large amounts of training data and may generalize poorly outside the training distribution. In this work, we present RIRFlow, a training-free Bayesian framework for RIR inverse problems using flow matching.
We derive a flow-consistent analytic prior from the statistical structure of RIRs, eliminating the need for data-driven priors.
Specifically, we model RIR as a Gaussian process with exponentially decaying variance, which yields a closed-form minimum mean squared error (MMSE) Wiener denoiser. This analytic denoiser is integrated as a prior in an existing flow-based inverse solver, where inverse problems are solved via guided posterior sampling. Furthermore, we extend the solver to nonlinear and non-Gaussian inverse problems via a local Gaussian approximation of the guided posterior, and empirically demonstrate that this approximation remains effective in practice.
Experiments on real RIRs across different inverse problems demonstrate robust performance, highlighting the effectiveness of combining a classic RIR model with the recent flow-based generative inference.
\end{abstract}


\maketitle

\section{Introduction}

Obtaining clean room impulse responses (RIR) may involve various  acoustic signal processing tasks, including denoising, deconvolution, inpainting, and declipping \cite{lin2006bayesian,crocco2015room, jalmby2023low, pezzoli2022deep, caviedes2023spatio, van2008optimally, lee2026anyrir, antonello2017room, sundstrom2024estimation, lemercier2025unsupervised}. These problems can be formulated as inverse problems of the form
\begin{align}
\mathbf{y} = \mathcal{H}(\mathbf{x}) + \mathbf{n},
\end{align}
where $\mathbf{x}$ denotes the unknown RIR, $\mathcal{H}$ is a linear or nonlinear forward operator, $\mathbf{n}$ represents additive noise, and $\mathbf{y}$ is the observed measurement. 
In practice, such inverse problems are often ill-posed due to non-invertible forward operators and measurement noise.
 
One common approach is to learn a direct mapping from measurements to RIRs using supervised learning. While effective in some settings, such methods require large paired datasets and often generalize poorly outside the training distribution \cite{richard2022deep, lluis2020sound, van2025deep, karakonstantis2024room}. A more principled alternative is Bayesian inference, which explicitly models prior knowledge about RIRs and estimates the posterior distribution \cite{xiang2020model}
\begin{align}
    p(\mathbf{x} \mid \mathbf{y}) \propto p(\mathbf{y} \mid \mathbf{x}) p(\mathbf{x}). \label{eq:basic_bayes}
\end{align}

Classical Bayesian approaches often rely on a single-point estimate of the posterior, most commonly the maximum a posteriori (MAP) point. In practice, this leads to optimization problems of the form
\begin{align}
    \min_{\mathbf{x}} \; -\log p(\mathbf{y} \mid \mathbf{x}) + \lambda \mathcal{R}(\mathbf{x}),
\end{align}
where $\mathcal{R}(\mathbf{x}) = -\log p(\mathbf{x})$ acts as a regularizer. For RIR estimation problems, common choices include $\ell_2$ (Tikhonov) regularization for Gaussian prior or $\ell_1$ for promoting sparsity, or handcrafted priors reflecting well-known properties of RIRs \cite{lin2006bayesian, van2008optimally, crocco2015room, jalmby2023low, pezzoli2022deep, caviedes2023spatio}. In the image domain, the use of regularizers has been extended through plug-and-play (PnP) \cite{venkatakrishnan2013plug, zhang2021plug} and regularization-by-denoising (RED) \cite{romano2017little} frameworks. 

While effective, such point-estimate methods do not capture posterior uncertainty inherent in ill-posed problems. In contrast, posterior sampling-based approaches aim to approximate the full posterior distribution and have been shown to better preserve perceptual fidelity \cite{blau2018perception}. Recent advances in diffusion and flow-based generative models enable such sampling in high-dimensional spaces by progressively transforming a simple distribution into a complex data distribution through a sequence of small, tractable steps \cite{sohl2015deep,song2020score,lipman2022flow}. A key advantage of this formulation is that inference proceeds through a sequence of incremental updates, each operating on a locally well-behaved distribution. This is particularly beneficial when the data distribution lies near a complex, low-dimensional manifold, where one-shot optimization methods such as MAP estimation becomes difficult and can lead to unstable results. 

In inverse problems, this framework enables efficient posterior sampling of $p(\mathbf{x}\mid\mathbf{y})$ while enforcing consistency with the observed measurements $\mathbf{y}$. Moreover, through Tweedie’s formula \cite{efron2011tweedie}, the score function is directly related to the minimum mean squared error (MMSE) denoiser, allowing denoisers to be interpreted as priors within diffusion- and flow-based inference frameworks \cite{song2019generative,song2020score}. This approach, often referred to as training-free guidance, does not require retraining the generative model for each inverse problem \cite{chung2022diffusion,ye2024tfg,pourya2025flower, pokle2023training, zhu2023denoising, martin2024pnp,song2023pseudoinverse}.

This paper introduces RIRFlow, which applies flow-based solvers for a range of RIR inverse problems using an analytic MMSE denoiser as a prior.  We derive this denoiser from statistical properties of RIRs. Rather than relying on data-driven denoisers, we show that a closed-form Wiener filter can be reformulated as the vector field of flow matching, yielding an interpretable and fully analytic prior. This formulation effectively bridges classical RIR modeling with modern flow-based generative inference. Building on the solver introduced in \cite{pourya2025flower}, we further extend the framework beyond its original setting. While the original method assumes linear forward models and Gaussian noise, we show that the approach remains applicable to nonlinear and non-Gaussian inverse problems by locally approximating the guided posterior with a Gaussian distribution. This approximation enables efficient inference via standard proximal optimization techniques while preserving the interpretation of posterior sampling.

Using this unified framework, we address denoising, $\ell_2$ and robust deconvolution, inpainting, and declipping.
Experiments demonstrate that the proposed method achieves stable and improved reconstructions across a wide range of inverse problems compared to baseline methods, particularly under low-SNR conditions.


This paper is structured as follows. Section~\ref{sec:background} introduces the relevant background for flow matching and flow-based solvers for inverse problems. Then, Section~\ref{sec:method} presents the analytic MMSE denoiser and how the existing flow-based solver can be extended for RIR inverse problems.  Section~\ref{sec:experiments} shows the experiments with results, and Section~\ref{sec:conclusion} summarizes the work.

\section{Background}\label{sec:background}





Flow matching has recently emerged as a unifying continuous-time generative modeling framework that generalizes diffusion models, offering a simpler and more efficient formulation while retaining their expressive power \cite{lipman2022flow}. The central idea of flow matching is to learn a vector field whose induced trajectories match a target probability flow. In this section, we review the fundamentals of flow matching and flow-based inverse solvers that form the foundation of the proposed method. 

\subsection{Flow matching}

Let $p_\text{src}$ and $p_\text{tgt}$ denote a source and target distribution, respectively. The target distribution is the data distribution that should be approximated with the model, a distribution of RIR, $p(\mathbf{x})$ in Eq.~\eqref{eq:basic_bayes}; the source distribution is a distribution that is easy to sample from, often, a standard normal distribution. Flow matching models the transformation between these distributions through a time-dependent probability density function, $p_{t}$, with $p_{0}(\mathbf{x})  = p_\text{src}(\mathbf{x})$ and $p_{1}(\mathbf{x}) = p_\text{tgt}(\mathbf{x})$. The flow is described via the ordinary differential equation (ODE): 
 \begin{align}
        \frac{d}{dt} \mathbf{x}_t = \mathbf{v}_t(\mathbf{x}_t), \quad t \in [0,1],
    \label{eq:ode}
 \end{align}
where $\mathbf{v}_t: \mathbb{R}^N  \rightarrow \mathbb{R}^N$ is a time-dependent vector field. The solution of this ODE induces a probability path $p_{t}$, such that the intermediate variable is distributed as $\textbf{X}_t \sim p_{t}$, with the goal that the endpoint satisfies $\textbf{X}_0 \sim p_{0}$ and $\textbf{X}_1 \sim p_{1}$.

The objective of flow matching is to learn a neural network $\mathbf{v}_t^\theta$, with parameters $\theta$, that approximates $\mathbf{v}_t$. This is achieved by minimizing the expected squared error
\begin{align}
L_{\mathrm{FM}}(\theta)
=
\mathbb{E}_{t \sim \mathcal{U}[0,1],\, \mathbf{X}_t \sim p_t}
\big[
\|\mathbf{v}_t^\theta(\mathbf{X}_t,t) - \mathbf{v}_t(\mathbf{X}_t)\|_2^2
\big].
\end{align}
Since the marginal vector field $\mathbf{v}_t(\mathbf{x}_t)$ is generally intractable, it is standard practice to instead consider the conditional vector field $\mathbf{v}_t(\mathbf{x}_t|\mathbf{x}_1)$. Under this conditioning, the probability path becomes tractable, and an effective choice is the straight-line flow, also known as the Gaussian conditional optimal transport probability path
\begin{align}
    \mathbf{X}_t = (1-t)\mathbf{X}_0 + t \mathbf{X}_1,
    \label{eq:straight-line-flow}
\end{align}
where \( \mathbf{X}_0 \sim \mathcal{N}(\mathbf{0}, \mathbf{I}) \) \cite{liu2022flow, holderrieth2025introduction}. This is also referred to as the forward process, which adds a given level of diffusion noise to the clean signal at each time step $t$. In this case of a straight path, it is simply a linear interpolation between $\mathbf{x}_0$ and $\mathbf{x}_1$. This implies that the conditional distribution is
\begin{align}
   p_t(\mathbf{x}_t \mid \mathbf{X}_1 = \mathbf{x}_1)
=
\mathcal{N}\!\left(\mathbf{x}_t;\, t\mathbf{x}_1, (1-t)^2 \mathbf{I}\right).
\end{align}
Then, the target vector field simplifies to $\mathbf{v}_t(\mathbf{x}_t \mid \mathbf{x}_1) = \mathbf{x}_1 - \mathbf{x}_0$ from Eq.~\eqref{eq:ode}, and the conditional flow matching (CFM) loss is
\begin{align}
L_{\mathrm{CFM}}(\theta)
=
\mathbb{E}_{t \sim \mathcal{U}[0,1],\, (\mathbf{X}_0,\mathbf{X}_1)\sim \pi}
\left[
\|\mathbf{v}_t^\theta(\mathbf{X}_t,t)
-
(\mathbf{X}_1 - \mathbf{X}_0)\|_2^2
\right],
\end{align}
where $\pi$ denotes a coupling between $p_0$ and $p_1$. A simple choice is the independent coupling
$\pi = p_0 \otimes p_1$, in which $\mathbf{X}_0$ and $\mathbf{X}_1$ are sampled independently.
The optimal solution of the CFM objective is given by
\begin{align}
\mathbf{v}_t^*(\mathbf{x}_t)
=
\mathbb{E}_{(\mathbf{X}_0,\mathbf{X}_1)\sim\pi}[\mathbf{X}_1 - \mathbf{X}_0 \mid \mathbf{X}_t = \mathbf{x}_t].
\label{eq:optimal_solution_CFM}
\end{align}

\subsection{Flow-based solver for inverse problems}
To avoid confusion with the notion of conditioning used in the derivation of the conditional flow matching loss above, conditioning on measurements is commonly referred to as \emph{guidance}. In this section, we briefly review how inverse problems can be addressed using training-free guidance within a flow-matching framework.
Flow-based solvers often rely on a pretrained, unguided flow matching model that defines a generative prior over the data distribution. While such a model allows sampling from the prior distribution $p_1(\mathbf{x})$, solving inverse problems requires sampling from the posterior distribution $p(\mathbf{x}_1 \mid \mathbf{Y} = \mathbf{y})$, which incorporates information from the measurements $\mathbf{y}$. Introducing this conditioning is non-trivial, since the flow model itself is not trained with measurements. While retraining the model for each inverse problem is possible, it requires paired data and limits generality. Recent work has therefore focused on such training-free guidance methods, which modify the sampling dynamics to enforce measurement consistency without retraining the model \cite{chung2022diffusion,ye2024tfg,pourya2025flower, pokle2023training, zhu2023denoising, martin2024pnp, song2023pseudoinverse}.


In this work, we adopt the FLOWER framework \cite{pourya2025flower}, which enables approximate posterior sampling. FLOWER follows a three-step procedure:
(i) denoising, which produces an initial estimate consistent with the prior distribution,  
(ii) measurement-aware refinement, where this estimate is adjusted using the observation model via a guidance step, and  
(iii) time progression, in which the refined estimate is mapped back onto the flow trajectory to advance the sampling process.

The FLOWER method focuses on a measurement operator $\mathcal{H}(\mathbf{x})$ that is linear and thus reduces to $\mathbf{H}\mathbf{x}$. The first step corresponds to a one-step denoising operation.
Given a noisy latent variable $\mathbf{X}_t$, the flow model produces an estimate of the
clean signal via
\begin{align}
\hat{\mathbf{x}}_1(\mathbf{x}_t)
=
\mathbb{E}[\mathbf{X}_1 \mid \mathbf{X}_t = \mathbf{x}_t]
=
\mathbf{x}_t + (1 - t)\mathbf{v}_t^*(\mathbf{x}_t),
\label{eq:denoiser}
\end{align}
which follows directly from the straight path in
Eq.~\eqref{eq:straight-line-flow} and the optimal vector field in
Eq.~\eqref{eq:optimal_solution_CFM}.

In the second step, measurement information is incorporated by approximating the guided distribution 
$p(\mathbf{x}_1|\mathbf{X}_t = \mathbf{x}_t, \mathbf{Y} = \mathbf{y})$.
To handle the confound between measurement noise $\mathbf{n}$ and diffusion noise, FLOWER adopts the $\Pi$GDM approximation \cite{song2023pseudoinverse}, in which the denoising distribution is approximated as
\begin{align}
 p(\mathbf{x}_1 \mid \mathbf{X}_t = \mathbf{x}_t)
\approx
\mathcal{N}\!\left(
\hat{\mathbf{x}}_1(\mathbf{x}_t),\,
\nu_t^2 \mathbf{I}
\right),
\label{eq:pigdm}
\end{align}
where the variance $\nu_t$ is annealed over time and satisfies
\(
\nu_t = \frac{1 - t}{\sqrt{t^2 + (1 - t)^2}},
\)
ensuring consistency with the marginal distribution $p_t$ as $t \to 1$. Assuming a linear forward operator with Gaussian measurement noise $p(\mathbf{y} \mid \mathbf{X}_1 = \mathbf{x}_1)
= \mathcal{N}(\mathbf{H}\mathbf{x}_1, \sigma_n^2 \mathbf{I})$, Pourya et al. \cite{pourya2025flower} show that $p(\mathbf{x}_1 \mid \mathbf{X}_t = \mathbf{x}_t, \mathbf{Y} = \mathbf{y})$ can be  approximated with a Gaussian distribution of the form
\begin{align}
    \tilde{p}(\mathbf{x}_1 \mid \mathbf{X}_t = \mathbf{x}_t,\, \mathbf{Y} = \mathbf{y})
= \mathcal{N}\!\left(\boldsymbol{\mu}_t(\mathbf{x}_t, \mathbf{y}), \boldsymbol{\Sigma}_t \right),
\label{eq:pigdm_flower_approx}
\end{align}

\noindent where 
\begin{align}
\boldsymbol{\mu}_t(\mathbf{x}_t, \mathbf{y})
&= \left( \nu_t^{-2} \mathbf{I} + \sigma_n^{-2} \mathbf{H}^\top \mathbf{H} \right)^{-1}
\left( \nu_t^{-2} \hat{\mathbf{x}}_1(\mathbf{x}_t)
      + \sigma_n^{-2} \mathbf{H}^\top \mathbf{y} \right), \label{eq:mu_t} \\
\boldsymbol{\Sigma}_t
&= \left( \nu_t^{-2} \mathbf{I} + \sigma_n^{-2} \mathbf{H}^\top \mathbf{H} \right)^{-1}.
\label{eq:sigma_t}
\end{align}

This Gaussian approximation provides a tractable surrogate for the true posterior and enables posterior sampling within the flow framework. In this view, FLOWER performs Bayesian inference by combining a flow-based prior with a measurement likelihood. Equivalently to Eq.~\eqref{eq:mu_t}, the update of the posterior mean can be written as a proximal operator
\begin{align}
\boldsymbol{\mu}_t(\mathbf{x}_t, \mathbf{y})
=
\operatorname{prox}_{\nu_t^2 F_{\mathbf{y}}}
\bigl(\hat{\mathbf{x}}_1(\mathbf{x}_t)\bigr),
\end{align}
where the proximal operator is defined as
\begin{align}
\operatorname{prox}_{\nu_t^2 F_{\mathbf{y}}}
\bigl(\hat{\mathbf{x}}_1(\mathbf{x}_t)\bigr)
=
\arg\min_{\mathbf{x}}
\left( F_{\mathbf{y}}(\mathbf{x})
+
\frac{1}{2\nu_t^2}
\|\mathbf{x}-\hat{\mathbf{x}}_1(\mathbf{x}_t)\|_2^2\right).
\label{eq:generic_prox}
\end{align}


As shown in \cite{pourya2025flower}, if $F_{\mathbf{y}}(\mathbf{x})
=
\frac{1}{2\sigma_n^2} \| \mathbf{H}\mathbf{x} - \mathbf{y} \|_2^2$, then the minimizer of the proximal gives Eq.~\eqref{eq:mu_t}.
The refined estimate is obtained via the reparameterization trick
\begin{align}
\begin{aligned}
\tilde{\mathbf{x}}_1(\mathbf{x}_t, \mathbf{y})
&= \boldsymbol{\mu}_t(\mathbf{x}_t, \mathbf{y}) + \gamma \boldsymbol{\kappa}_t, \\
\boldsymbol{\kappa}_t &\sim \mathcal{N}(0, \boldsymbol{\Sigma}_t),
\label{eq:reparametrization_trick}
\end{aligned}
\end{align}
where setting the uncertainty $\gamma = 1$ corresponds to exact posterior sampling. Sampling $\boldsymbol{\kappa}_t$ is done by 
\begin{align}
    \boldsymbol{\kappa}_t = \boldsymbol{\Sigma}_t (\nu_t^{-1}\boldsymbol{\varepsilon}_1 + \sigma_n^{-1}\mathbf{H}^T\boldsymbol{\varepsilon}_2),
\end{align}
where $\boldsymbol{\varepsilon}_1, \boldsymbol{\varepsilon}_2 \sim \mathcal{N}(0, \mathbf{I})$.

Finally, the refined estimate is projected back onto the flow path
\begin{align}
\mathbf{x}_{t+\Delta t}
=
(1 - t - \Delta t)\boldsymbol{\epsilon}
+
(t + \Delta t)\tilde{\mathbf{x}}_1(\mathbf{x}_t, \mathbf{y}),
\end{align}
with $\epsilon \sim \mathcal{N}(0, \mathbf{I})$, and the procedure is repeated T time steps to yield $\mathbf{x}_1$.
It is important to emphasize that this is a posterior sampling approach, which implies that there is an infinite set of possible solutions. As a result, chosing a strategy to select one of the solutions is another design choice.


Many existing inverse-problem solvers are limited to linear forward models and Gaussian measurement noise, with only a few methods addressing nonlinear or non-Gaussian settings \cite{song2023pseudoinverse, kawar2022jpeg}. This limitation arises because, in such cases, the Gaussian assumption underlying the guided distribution 
$p(\mathbf{x}_1 \mid \mathbf{X}_t = \mathbf{x}_t, \mathbf{Y} = \mathbf{y})$
no longer holds. FLOWER is also derived under the assumption of linear measurement operators and Gaussian noise. However, its formulation naturally allows extensions by modifying the refinement step of the algorithm, which incorporates measurement information. Although the original work does not present empirical results for nonlinear inverse problems, it is noted that such cases could, in principle, be handled by replacing the closed-form proximal update with a sampling-based procedure, such as Langevin dynamics using the score function.



\subsection{vector field and MMSE denoiser}
Once an optimal vector field $\mathbf{v}_t^*$ has been learned through flow matching, a denoiser naturally emerges as a byproduct of the formulation.
As shown in Eq.~\eqref{eq:denoiser}, the MMSE denoiser is given by \cite{martin2024pnp, gagneux2025generation}
\begin{align}
\begin{aligned}
    D_t(\mathbf{x}_t)
    &= \mathbb{E}[\mathbf{X}_1 \mid \mathbf{X}_t = \mathbf{x}_t],
\end{aligned} \label{eq:denoiser_from_flow}
\end{align}
which corresponds to the optimal estimator of the clean signal \( \mathbf{x}_1 \) given the noisy observation \( \mathbf{x}_t \). Conversely, this relationship implies that an MMSE denoiser determines the corresponding vector field. An optimal vector field can be obtained as 
\begin{align}
\begin{aligned}
    \mathbf{v}^*(\mathbf{x}_t, t)
    =
    \frac{D_t(\mathbf{x}_t) - \mathbf{x}_t}{1 - t}.
    \label{eq:vec_from_denoiser}
\end{aligned}
\end{align}
In this work, we exploit this duality by computing the vector field directly from an analytic MMSE denoiser. This allows us to employ a flow-based solver without learning a neural-network-based vector field, while still retaining the benefits of flow-based inference. As a result, the proposed method leverages the structure of the flow model while the prior remains fully analytic and interpretable.

\section{Method}\label{sec:method}
In this section, we show the derivation of the MMSE denoiser for RIRs that aligns with flow matching.  Then, we present how FLOWER is applied to five different RIR inverse problems, with both linear and nonlinear forward operators and Gaussian and non-Gaussian measurement noises.

\subsection{Analytic MMSE denoiser} \label{subsection:MMSE_denoiser}




RIRs can be well approximated as exponentially decaying Gaussian noise \cite{Kuttruff2009RA}. While this simple model does not explicitly represent the direct sound or early reflections, it remains well-suited as a prior for inverse problems. In practice, the direct path and early reflections typically exhibit a high SNR, so the prior has little influence in these regions compared to the observed measurement signal. This naturally results from a precision-weighted trade-off between likelihood and prior.

Under the assumption of Gaussian noise with exponentially decaying variance, the MMSE estimator admits a closed-form solution and is equivalent to a Wiener filter. If the measurement noise is not necessarily Gaussian, the denoiser is no longer optimal in an MMSE sense. Nevertheless, it can remain an effective denoiser.

In the present setting, the Wiener filter depends on the signal variance, which is governed by the decay rate and initial amplitude of the RIR. Since these parameters are generally unknown, they must be inferred from the observation itself.
We address this by estimating the decay parameters in a Bayesian manner, yielding an MMSE Wiener filter. The resulting estimator is fully consistent with the flow matching formulation and provides an analytic MMSE denoiser that can be directly embedded into the flow-based inference framework.

Let $\mathbf{X}_1 \rightarrow \mathbb{R}^N$ denote the random variable corresponding to a clean RIR, 
$\mathbf{X}_0 \rightarrow \mathbb{R}^N$ pure Gaussian noise, and $\mathbf{X}_t \rightarrow \mathbb{R}^N$ an intermediate state at time $t$ in the flow-matching process, with noise variance $(1-t)^2$. This notation is consistent with the formulation of flow matching used throughout this work.

We model each time sample $k$ of the RIR, $X_{1,k} \rightarrow \mathbb{R}$ as a zero-mean white Gaussian variable with exponentially decaying variance,
\begin{align}
X_{1,k} \sim \mathcal{N}\!\left(0,\; d_k(\lambda,\alpha)\right),
\qquad
d_k(\lambda,\alpha) = \alpha^2 e^{-2\lambda k},
\label{eq:rir_model}
\end{align}
where $\lambda$ controls the decay rate and $\alpha$ determines the initial amplitude.
Stacking all time samples into a vector $\mathbf{X}_1$, the RIR distribution can be written compactly as
\begin{align}
\mathbf{X}_1 \sim \mathcal{N}\!\left(\mathbf{0}, \mathbf{D}(\lambda,\alpha)\right),
\end{align}
where $\mathbf{D}(\lambda,\alpha)$ is a diagonal covariance matrix whose entries are given by $d_k(\lambda,\alpha)$. 

The straight path in Eq.~\eqref{eq:straight-line-flow} can also be written as 
\begin{align}
\mathbf{X}_{t} = t \mathbf{X}_{1} + (1-t)\boldsymbol{\varepsilon},
\qquad \boldsymbol{\varepsilon} \sim \mathcal{N}(\mathbf{0},\mathbf{I}).
\end{align}
Since $\mathbf{X}_{1}$ and $\boldsymbol{\varepsilon}$ are independent Gaussian variables, $\mathbf{X}_{t}$ is also Gaussian with variance
\begin{align}
\mathrm{Var}(\mathbf{X}_{t})
= t^2 \mathbf{D}(\lambda,\alpha) + (1-t)^2,
\end{align}
yielding the following distribution for $\mathbf{X}_{t}$
\begin{align}
p(\mathbf{x}_{t}\mid\lambda,\alpha) =
\mathcal{N}\!\left(
0,\;
t^2 \mathbf{D}(\lambda,\alpha) + (1-t)^2 \mathbf{I}
\right).
\label{eq:likelihood_model_x_t}
\end{align}

The combination of a Gaussian prior and a linear Gaussian forward process implies that the conditional distribution
$p(\mathbf{x}_1 \mid \mathbf{X}_t=\mathbf{x}_t, \lambda,\alpha)$ remains Gaussian for all $t\in[0,1]$.
Consequently, the MMSE denoiser admits the closed form
\begin{align}
\begin{aligned}
\mathbb{E}[\mathbf{X}_{1}\mid \mathbf{X}_{t}=\mathbf{x}_{t},\lambda,\alpha]
&=
\frac{\mathrm{Cov}(\mathbf{X}_{1},\mathbf{X}_{t})}{\mathrm{Var}(\mathbf{X}_{t})}\odot\,\mathbf{x}_{t} \\
&=
\frac{t\,\mathbf{D}(\lambda,\alpha)}{t^2 \mathbf{D}(\lambda,\alpha) + (1-t)^2\mathbf{I}}
\odot\,\mathbf{x}_{t}\\
&\;\triangleq\;
\tilde{\boldsymbol{\beta}}(\lambda,\alpha,t)\,\mathbf{x}_{t}.
\label{eq:mmse_denoiser}
\end{aligned}
\end{align}
Since $\mathbf{D}(\lambda,\alpha)$ is diagonal, this estimator acts elementwise. In scalar form,
\begin{align}
\mathbb{E}[X_{1,k} \mid X_{t,k}=x_{t,k},\lambda,\alpha]
=
\tilde{\beta}_k(\lambda,\alpha,t)\,x_{t,k},
\end{align}
with the shrinkage factor 
\begin{align}
\tilde{\beta}_k(\lambda,\alpha,t)
=
\frac{t\,d_k(\lambda,\alpha)}
     {t^2 d_k(\lambda,\alpha)+(1-t)^2}.
     \label{eq:shrinkage_factor}
\end{align}

This expression admits a clear interpretation as a Wiener filter, where for every time sample $k$,
$t^2 d_k(\lambda,\alpha)$ corresponds to the signal variance and $(1-t)^2$ to the noise variance.
Consequently, this denoiser smoothly interpolates between pure noise removal at early diffusion
times ($t \to 0$) and identity mapping as $t \to 1$, consistent with the behavior of flow matching generative models.

Since the parameters $\lambda$ and $\alpha$ are unknown, we marginalize them out using the law of total expectation:
\begin{align}
\begin{aligned}
\mathbb{E}[\mathbf{X}_1 \mid \mathbf{X}_t = \mathbf{x}_t]
&=
\mathbb{E}_{\lambda, \alpha \mid \mathbf{X}_t = \mathbf{x}_t}
\!\left[
\mathbb{E}[\mathbf{X}_1 \mid \mathbf{X}_t = \mathbf{x}_t, \lambda, \alpha]
\right] \\
&=
\mathbb{E}_{\lambda, \alpha \mid \mathbf{X}_t = \mathbf{x}_t}
\!\left[
\tilde{\boldsymbol{\beta}}(\lambda,\alpha)\,\mathbf{x}_t
\right].
\end{aligned}
\end{align}

The posterior over the decay parameters follows from Bayes’ rule,
\begin{align}
p(\lambda, \alpha \mid \mathbf{X}_t = \mathbf{x}_t)
\propto
p(\mathbf{x}_t \mid \lambda, \alpha)\,p(\alpha)\,p(\lambda).
\label{eq:decay_param_bayes}
\end{align}

To avoid joint integration over $(\lambda,\alpha)$, we treat $\lambda$ as the parameter of interest and profile out $\alpha$, following the profile likelihood approach \cite{kreutz2013profile}.
For a fixed decay rate $\lambda$, the optimal value of $\alpha$ admits a closed-form solution based on the energy decay curve (EDC).

From the exponential RIR model in Eq.~\eqref{eq:rir_model}, the expected theoretical EDC at time index $m$ is
\begin{align}
\begin{aligned}
\mathbb{E}[\mathrm{EDC}_t(m)]
&=
\sum_{k=m}^{N-1}
\mathbb{E}[X_{t,k}^2] \\
&=
\sum_{k=m}^{N-1}
\left(
t^2 \alpha^2 e^{-2\lambda k}
+
(1-t)^2
\right) \\
&=
t^2 \alpha^2 \psi_\lambda(m)
+
(1-t)^2 (N-m),
\end{aligned}
\end{align}
where $\psi_\lambda(m) = \sum_{k=m}^{N-1} e^{-2\lambda k}$
is a deterministic basis function of $\lambda$.

Meanwhile, given an observation $\mathbf{x}_t$, the empirical EDC is
\begin{align}
\mathrm{EDC}_{\mathrm{obs}}(m)
=
\sum_{k=m}^{N-1} x_{t,k}^2,
\end{align}
and we can say that the noise-compensated EDC is
\begin{align}
\mathrm{EDC}_{\mathrm{sig}}(m)
=
\mathrm{EDC}_{\mathrm{obs}}(m)
-
(1-t)^2 (N-m).
\end{align}


For a fixed decay rate $\lambda$, the optimal $\alpha$ is obtained by solving
\begin{align}
\hat{\alpha}^2(\lambda)
=
\arg\min_s
\sum_m
\left(
\mathrm{EDC}_{\mathrm{sig}}(m)
-
t^2 s\,\psi_\lambda(m)
\right)^2,
\end{align}
which has the closed-form solution
\begin{align}
\hat{\alpha}^2(\lambda)
=
\frac{\sum_m \psi_\lambda(m)\,\mathrm{EDC}_{\mathrm{sig}}(m)}
     {t^2\sum_m \psi_\lambda(m)^2}.
\end{align}

Substituting $\hat{\alpha}(\lambda)$ back into the model (Eq.~\eqref{eq:rir_model}) yields the $\alpha$-profiled variance 
\begin{align}
\hat{d}_k(\lambda)
=
\hat{\alpha}^2(\lambda)\,e^{-2\lambda k}.
\end{align}

Then the profile likelihood is
\begin{align}
p(\mathbf{x}_t\mid\lambda)
&=
\prod_{k=0}^{N-1}
\mathcal{N}\!\left(
x_{t,k};\,0,\; t^2 \hat{d}_k(\lambda) + (1-t)^2
\right) 
\label{eq:profile_likelihood}
\end{align}

We treat the decay rate $\lambda$ as a latent hyperparameter and marginalize it using a grid-based Bayesian model averaging scheme.
Specifically, $\lambda$ is parameterized in terms of the reverberation time $\mathrm{RT}$ as
\begin{align}
    \lambda = \frac{6.90}{\mathrm{RT}\,f_\mathrm{s}},
\end{align}
where $f_\mathrm{s}$ is the sampling rate. The parameter $\lambda$ is discretized over a finite range corresponding to plausible reverberation times
$\mathrm{RT} \in [0.1,\, 3.0]$.

The final analytic MMSE denoiser is obtained by averaging over the posterior $p(\lambda \mid \mathbf{x}_t)$,
\begin{align}
D_{t,k}(\mathbf{x}_t)
= \mathbb{E}[X_{1,k} \mid \mathbf{X}_t=\mathbf{x}_t] = 
\sum_{\ell}
\omega_\ell(\mathbf{x}_t)\,
\beta_k(\lambda_\ell,t)\,x_{t,k},
\end{align}
where $\omega_\ell(\mathbf{x}_t) \propto p(\mathbf{x}_t \mid \lambda_\ell)\,p(\lambda_\ell)$ and the shrinkage factor in Eq.~\eqref{eq:shrinkage_factor} simplified to 
\begin{align}
\beta_k(\lambda_\ell,t)
=
\frac{t\,\hat{d}_k(\lambda_\ell)}
{t^2\,\hat{d}_k(\lambda_\ell)+(1-t)^2}.
\end{align}

Now we can use Eq.~\eqref{eq:vec_from_denoiser} to
obtain the vector field and will act as a prior in the flow-based solver.    


So far, we have derived the MMSE denoiser under a single-band RIR model. 
In practice, however, RIRs have frequency-dependent decay rates.
To account for this, we extend the formulation to an octave-band model by assuming that the decay rates in different frequency bands are statistically independent. Under this assumption, the full-band RIR distribution is obtained by applying the single-band MMSE formulation independently to each octave band. 
Consequently, the multi-band denoiser is constructed by computing the MMSE estimate in each band separately and combining the results, yielding a joint posterior that captures frequency-dependent reverberation characteristics while retaining the analytical tractability of the single-band model.





\begin{algorithm}[t]
\caption{RIRFlow}
\label{alg:rirflow}
\begin{algorithmic}[1]
\REQUIRE $\mathbf{y}$, forward operator $\mathbf{H}$, noise level $\sigma_n$, MMSE denoiser $D_t(\cdot)$, steps $T$, uncertainty $\gamma\in[0,1]$, band split/merge $\mathcal{B},\mathcal{S}$, bands $B$.
\STATE Sample $\mathbf{x}_0 \sim p_{\mathbf{x}_0}$.
\FOR{$j=0$ to $T-1$}
    \STATE $t \gets \dfrac{1-\cos(\pi j/T)}{2}$
    \STATE $\nu_t \gets \dfrac{1-t}{\sqrt{t^2+(1-t)^2}} \;$
    \STATE \textbf{(Step 1) Denoiser:}
    \STATE $\{\mathbf{x}_t^{(b)}\}_{b=1}^B \gets \mathcal{B}(\mathbf{x}_t)$
    \FOR{$b=1$ to $B$}
        \STATE $\hat{\mathbf{x}}_1^{(b)} \gets D_t(\mathbf{x}_t^{(b)})$.
        \STATE $\mathbf{v}_t^{(b)} \gets \dfrac{\hat{\mathbf{x}}_1^{(b)}-\mathbf{x}_t^{(b)}}{1-t}$
    \ENDFOR
    \STATE $\hat{\mathbf{x}}_1 \gets \mathcal{S}(\{\hat{\mathbf{x}}_1^{(b)}\}_{b=1}^B)$
    \STATE \textbf{(Step 2a) Measurement-aware Refinement:}
    \STATE $\boldsymbol{\mu}_t \gets \arg\min_{\mathbf{x}}\;F_{\mathbf{y}}(\mathbf{x})+\dfrac{1}{2\nu_t^2}\|\mathbf{x}-\hat{\mathbf{x}}_1\|_2^2$
    \STATE \textbf{(Step 2b) Uncertainty:} sample $\boldsymbol{\varepsilon}_1,\boldsymbol{\varepsilon}_2\!\sim\!\mathcal{N}(\mathbf{0},\mathbf{I})$, 
    \STATE $\boldsymbol{\kappa}_t =    \boldsymbol{\Sigma}_t(\nu_t^{-1}\boldsymbol{\varepsilon}_1+\sigma_n^{-1}\mathbf{H}^T\boldsymbol{\varepsilon}_2)$
    \STATE $\tilde{\mathbf{x}}_1 \gets \mu_t + \gamma\,\boldsymbol{\kappa}_t$.
    \STATE \textbf{(Step 3) Time progression:} sample $\boldsymbol{\epsilon} \sim \mathcal{N}(\mathbf{0},\mathbf{I})$,
    \STATE $\mathbf{x}_{t+\Delta t} \gets (1-t-\Delta t)\boldsymbol{\epsilon} + (t+\Delta t)\tilde{\mathbf{x}}_1$

\ENDFOR

\Return $\mathbf{x}_1 \gets \tilde{\mathbf{x}}_1$
\end{algorithmic}
\end{algorithm}

\subsection{Applications to inverse problems}

\begin{figure*}[t!]
\centering
\includegraphics[]{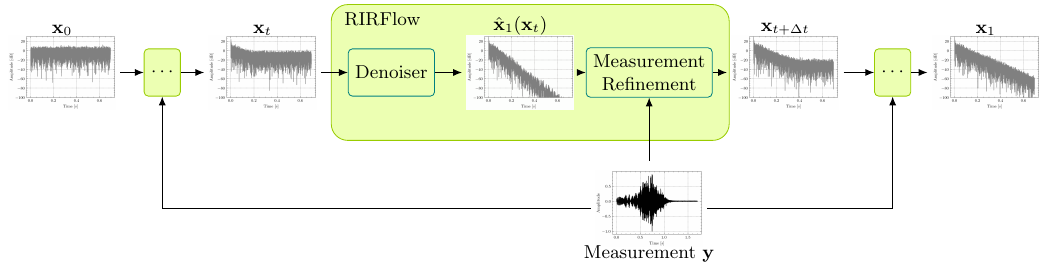}
\caption{Illustration of RIRFlow. The algorithm starts from an initial noise sample $\mathbf{x}_0 \sim p_0$ and iteratively applies denoising followed by measurement refinement. Through this process, it yields a sample $\mathbf{x}_1 \sim p_1$ that is consistent with both the RIR prior and the observed measurement.}
\label{fig:flow_diagram}
\end{figure*}

Having derived an analytic MMSE denoiser that serves as a generative prior for RIRs, we now apply it to a range of inverse problems within the FLOWER framework and refer to the resulting method as RIRFlow. The key idea is to combine the analytic prior with a task-specific measurement model, enabling a unified treatment of different RIR inverse problems. An overview of the algorithm is illustrated in Fig.~\ref{fig:flow_diagram} and summarized in Algorithm~\ref{alg:rirflow}

All inverse problems considered in this work are solved using the same RIRFlow pipeline; they differ only in how the measurement-consistency step is formulated. Given a noisy intermediate estimate $\mathbf{x}_t$, the first step applies the analytic MMSE denoiser to obtain a denoised estimate $\hat{\mathbf{x}}_1(\mathbf{x}_t)$ that lies on the prior distribution $p_1$. The denoiser is applied independently in each octave band, after which the bandwise estimates are merged to form a full-band RIR estimate.

The second step enforces consistency with the measurement by solving the proximal problem in Eq.~\eqref{eq:generic_prox} using a task-specific data-fidelity term $F_{\mathbf{y}}(\mathbf{x})$. This step is common to all inverse problems considered in this work; only the form of $F_{\mathbf{y}}(\mathbf{x})$ changes depending on the forward model and noise assumptions.

In the original FLOWER formulation \cite{pourya2025flower}, the guided distribution
$\tilde{p}(\mathbf{x}_1 \mid \mathbf{X}_t=\mathbf{x}_t, \mathbf{Y}=\mathbf{y})$
is Gaussian, which allows exact sampling. This follows from the assumption of a linear forward model and Gaussian measurement noise. Under more general conditions, such as non-Gaussian noise or nonlinear forward models, this distribution is no longer Gaussian and closed-form sampling is no longer available. 
Nevertheless, we show experimentally that the framework remains applicable by replacing the exact guided distribution with a local Gaussian approximation. In particular, the guided distribution is approximated using its first- and second-order moments, obtained via a second-order expansion of the data-fidelity term. This yields a tractable Gaussian approximation that preserves the interpretation of posterior sampling in FLOWER while enabling efficient optimization using standard proximal methods \cite{parikh2014proximal}. The third step then follows the original FLOWER formulation \cite{pourya2025flower}, advancing the estimate along the flow trajectory.

In the following subsections, we formulate several RIR inverse problems and demonstrate how they can be solved using RIRFlow. While the overall algorithm remains the same, the forward model and noise assumptions lead to different realizations of the measurement refinement step.

\subsubsection{Denoising}\label{sec:denoising}
As RIRs are obtained through real-world measurements, background noise is unavoidable and always a nuisance in obtaining a clean RIR. Therefore, denoising is an important post-processing step in measurement systems to obtain a clean RIR \cite{goodwin2006analysis, cabrera2011decay, masse2020robust, hold2022resynthesis, prawda2025cropping}. The denoising inverse problem is the simplest case, as the forward operator is the identity, i.e., \(\mathbf{H}=\mathbf{I}\). Thus it is formulated as 
\begin{align}
\begin{aligned}
\mathbf{y} &= \mathbf{x} + \mathbf{n},\\
\mathbf{n} &\sim \mathcal{N}(\mathbf{0}, \sigma_n^2 \mathbf{I}),
\end{aligned}
\end{align}
where $\mathbf{x} \in \mathbb{R}^N$ is the clean RIR and $\mathbf{n} \in \mathbb{R}^N$ is Gaussian background noise and $\mathbf{y} \in \mathbb{R}^N$ is the noisy measured RIR. 
The measurement-consistent refinement step to compute the guided flow corresponds to solving the following proximal
\begin{align}
\boldsymbol{\mu}_t(\mathbf{x}_t, \mathbf{y})
&= \arg\min_{\mathbf{x}\in\mathbb{R}^N}
\underbrace{
\dfrac{1}{2\sigma_n^2}\|\mathbf{x} - \mathbf{y}\|_2^2
}_{F_\mathbf{y}(\mathbf{x})}
+
\frac{1}{2\nu_t^2}\|\mathbf{x}-\hat{\mathbf{x}}_1(\mathbf{x}_t)\|_2^2.
\end{align}
This quadratic objective has a closed-form solution 
\begin{align}
    \boldsymbol{\mu}_t(\mathbf{x}_t, \mathbf{y}) = \frac{\sigma_n^{-2}\mathbf{y}+\nu_t^{-2}\hat{\mathbf{x}}_1(\mathbf{x}_t)}{\sigma_n^{-2}+\nu_t^{-2}}, 
\end{align}
which corresponds to a precision-weighted average between the noisy observation
and the flow-based prior estimate.
The weights are determined by the measurement noise variance $\sigma_n^2$
and the time-dependent uncertainty $\nu_t^2$ induced by the flow.

The associated conditional covariance is
\begin{align}
\boldsymbol{\Sigma}_t
=
(\sigma_n^{-2}+\nu_t^{-2})^{-1}\mathbf{I},
\end{align}
and sampling is performed according to 
Eq.~\eqref{eq:reparametrization_trick}.

\subsubsection{$\ell_2$ Deconvolution}\label{sec:l2_deconvolution}

Deconvolution is a widely used inverse problem in room acoustic measurement systems for estimating an unknown RIR from a measured signal. In this work, we consider a non-blind deconvolution, where the excitation signal (e.g., an exponential sine sweep) is known. 
Classical approaches to deconvolution include spectral division \cite{farina_simultaneous_2000, muller_transfer-function_2001} or regularized least squares methods \cite{lin2006bayesian, sundstrom2024estimation, crocco2015room, jalmby2023low}.
Typically, measurements are assumed to be corrupted by additive, stationary Gaussian noise. Under this assumption, the $\ell_2$ deconvolution problem is formulated as
\begin{align}
\begin{aligned}
\mathbf{y} &= \mathbf{H}\mathbf{x} + \mathbf{n},\\
\mathbf{n} &\sim \mathcal{N}(\mathbf{0}, \sigma_n^2 \mathbf{I}),
\end{aligned}
\end{align}
where $\mathbf{x}\in\mathbb{R}^N$ is the unknown RIR, $\mathbf{H}$ denotes a convolution matrix with the known excitation signal $\mathbf{h}\in\mathbb{R}^M$, and $\mathbf{y}\in\mathbb{R}^{N+M-1}$ is the observed signal. In practice, $\mathbf{H}$ is not formed explicitly; instead, convolution and its adjoint are implemented efficiently via Fast Fourier Transforms (FFTs).

Given the current flow estimate $\hat{\mathbf{x}}_1(\mathbf{x}_t)$, the measurement-aware refinement step is
\begin{align}
\boldsymbol{\mu}_t(\mathbf{x}_t,\mathbf{y})
&= \arg\min_{\mathbf{x}}
\underbrace{\frac{1}{2\sigma_n^2}\|\mathbf{H}\mathbf{x}-\mathbf{y}\|_2^2}_{F_{\mathbf{y}}(\mathbf{x})}
+ \frac{1}{2\nu_t^2}\|\mathbf{x}-\hat{\mathbf{x}}_1(\mathbf{x}_t)\|_2^2,
\label{eq:l2_proximal}
\end{align}
which permits a closed-form solution as in Eq.~\eqref{eq:mu_t} and the covariance $\boldsymbol{\Sigma}_t$ is the same as in Eq.~\eqref{eq:sigma_t}.

Although a closed-form solution exists, explicitly forming or inverting the forward operator $\mathbf{H}$ is infeasible in practice.
Instead, the solution is obtained using conjugate gradients (CG), which only requires matrix–vector products.
Specifically, the refinement step reduces to
\begin{align}
\boldsymbol{\Sigma}_t^{-1} \boldsymbol{\mu}_t
&=
\nu_t^{-2}\hat{\mathbf{x}}_1(\mathbf{x}_t)
+
\sigma_n^{-2}\mathbf{H}^\top\mathbf{y},
\\
\boldsymbol{\Sigma}_t^{-1} \boldsymbol{\kappa}_t
&=
\nu_t^{-1}\boldsymbol{\varepsilon}_1
+
\sigma_n^{-1}\mathbf{H}^\top\boldsymbol{\varepsilon}_2,
\end{align}
where
$\boldsymbol{\varepsilon}_1,\boldsymbol{\varepsilon}_2 \sim \mathcal{N}(\mathbf{0},\mathbf{I})$.
Both systems are solved using CG with matrix-free FFT-based convolutions for $\mathbf{H}\mathbf{x} = \mathbf{h} * \mathbf{x}$ and $\mathbf{H}^\top \mathbf{x} = \tilde{\mathbf{h}} * \mathbf{x}$,
with $\tilde{\mathbf{h}}$ denoting the time-reversed filter.

\subsubsection{Robust deconvolution}\label{sec:l1_deconvolution}
Real-world acoustic measurements often contain not only stationary background noise, but also non-stationary and impulsive disturbances, such as door slams or interfering speech. While stationary noise can be reasonably modeled as Gaussian, such impulsive components exhibit heavy-tailed statistics, for which the Laplacian distribution is a common approximation. In a maximum likelihood sense, the squared $\ell_2$-norm is optimal under Gaussian noise assumptions, whereas the $\ell_1$-norm is optimal for Laplacian noise. Owing to its reduced sensitivity to outliers, the $\ell_1$-norm is therefore widely used as a robust estimator \cite{boyd2004convex}. In previous works, robust deconvolution significantly improves RIR estimation in the presence of real-world non-stationary noise~\cite{cang2022robust, lee2026anyrir}.

We consider the measurement model 
\begin{align}
\begin{aligned}
\mathbf{y} &= \mathbf{H}\mathbf{x} + \boldsymbol{\eta} + \mathbf{n}\\
\eta_i &\sim \mathcal{L} (0, b), \quad i=1,\dots,N\\
\mathbf{n} &\sim \mathcal{N}(\mathbf{0}, \sigma_n^2 \mathbf{I}),
\end{aligned}
\end{align}
where $\boldsymbol{\eta}$  represents Laplacian noise and $\mathbf{n}$ denotes Gaussian noise.

Under non-Gaussian measurement noise, the guided distribution $\tilde{p}(\mathbf{x}_1 \mid \mathbf{X}_t=\mathbf{x}_t, \mathbf{Y}=\mathbf{y})$ is no longer Gaussian. As aforementioned, our experiments show that the FLOWER framework remains valid by replacing it with a tractable approximation.
To account for both Gaussian and Laplacian noise components, we adopt a Huber loss, which combines the quadratic and linear penalties. The resulting proximal update is with the residual 
\begin{align}
    \mathbf{r} = \mathbf{H}\mathbf{x} - \mathbf{y},
\end{align}

\begin{align}
\mu_t(\mathbf{x}_t,\mathbf{y})
=
\arg\min_{\mathbf{x}}
\underbrace{
\frac{1}{\sigma_n^2}
\sum_i \rho_\delta(r_i)
}_{F_{\mathbf{y}}(\mathbf{x})}
+
\frac{1}{2\nu_t^2}
\|\mathbf{x}-\hat{\mathbf{x}}_1(\mathbf{x}_t)\|_2^2,
\label{eq:huber_prox}
\end{align}

where
\begin{align}
\rho_\delta(r)=
\begin{cases}
\frac{1}{2}r^2, & |r|\le\delta,\\
\delta\left(|r|-\frac{\delta}{2}\right), & |r|>\delta,
\end{cases}
\qquad
\delta = \sigma_n^2 / b .
\end{align}

The objective in Eq.~\eqref{eq:huber_prox} is minimized using iterative reweighted least squares (IRLS) with a quadratic form,
\begin{align}
\mu_t(\mathbf{x}_t,\mathbf{y})
=
\arg\min_{\mathbf{x}}
\frac{1}{2}\mathbf{r}^\top \mathbf{W}_{\mathrm{Huber}}\mathbf{r}
+
\frac{1}{2\nu_t^2}
\|\mathbf{x}-\hat{\mathbf{x}}_1(\mathbf{x}_t)\|_2^2,
\end{align}
with diagonal weights
\begin{align}
(\mathbf{W}_{\mathrm{Huber}})_{ii} =
\begin{cases}
\sigma_n^{-2}, & |r_i| \le \delta,\\[4pt]
\dfrac{1}{b|r_i|}, & |r_i| > \delta.
\end{cases}
\end{align}
Obtaining $\boldsymbol{\mu}_t$ and $\boldsymbol{\kappa}_t$  reduces to solving the linear system with CG as in $\ell_2$ deconvolution, where
\begin{align}
\boldsymbol{\Sigma_t}^{-1}
=
\nu_t^{-2}\mathbf{I}
+
\mathbf{H}^\top \mathbf{W}_{\mathrm{Huber}}\mathbf{H}.
\end{align}

\subsubsection{Inpainting}
\begin{figure}[t!]
\centering
\includegraphics[width=\columnwidth]{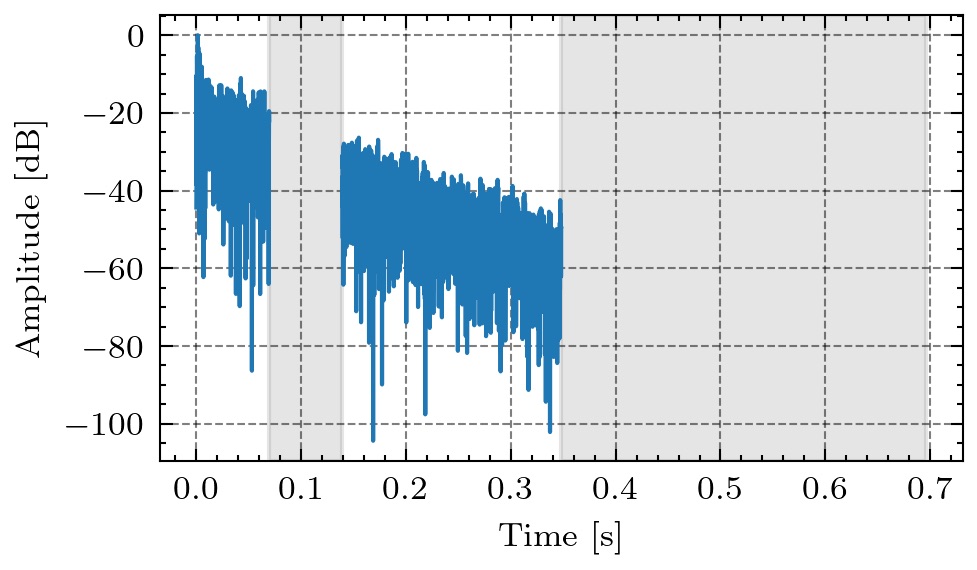}
\caption{Example of a cropped RIR (``h054\_Kitchen\_3txts'') for an inpainting problem.}
\label{fig:inpainting_example}
\end{figure}

As opposed to capturing noise during measurements with sweeps or other measurement signals, RIRs themselves may be corrupted by noise and interference. This occurs, for example, when performing balloon pop \cite{patynen2011investigations}, or hand-clap \cite{papadakis_handclap_2020} measurements in public spaces. Therein, RIRs are measured directly, and larger segments can be corrupted by noise in the background. The noisy segments can be cropped out as shown in Fig.~\ref{fig:inpainting_example}. The goal of inpainting for RIR is to resynthesize the corrupted segments in a manner that is as consistent with the non-corrupted segments. This is analogous to the inpainting problem in the image domain. 

The observation model is
\begin{align}
\begin{aligned}
\mathbf{y} &= \mathbf{M}\mathbf{x} + \mathbf{n},\\
\mathbf{n} &\sim \mathcal{N}(\mathbf{0},\sigma_n^2\mathbf{I}),
\end{aligned}
\end{align}
where \(\mathbf{x}\in\mathbb{R}^N\) is the clean RIR, \(\mathbf{y}\in\mathbb{R}^N\) is the corrupted measurement, and diagonal masking matrix 
\(\mathbf{M}\in\mathbb{R}^{NxN}\) where (\(M_{ii}=1\)) indicate observed and (\(M_{ii}=0\)) indicate missing samples.



This objective is quadratic and thus, gives the closed-form solution of $\boldsymbol{\mu}_t$ and $\mathbf{\Sigma}_t$ as in  Eq.~\eqref{eq:mu_t} and Eq.~\eqref{eq:sigma_t} with $\mathbf{H}=\mathbf{M}$.
Since \(\mathbf{M}\) is diagonal, it reduces to an elementwise estimation:
\begin{align}
\label{eq:inpainting_elementwise}
(\mu_t)_i
=
\begin{cases}
\dfrac{\sigma_n^{-2}y_i+\nu_t^{-2}\hat{x}_{1,i}(\mathbf{x}_t)}{\sigma_n^{-2}+\nu_t^{-2}},
& M_{ii}=1,\\[10pt]
\hat{x}_{1,i}(\mathbf{x}_t), & M_{ii}=0.
\end{cases}
\end{align}

\subsubsection{Clipped deconvolution}
\begin{figure}[t!]
\centering
\includegraphics[width=\columnwidth]{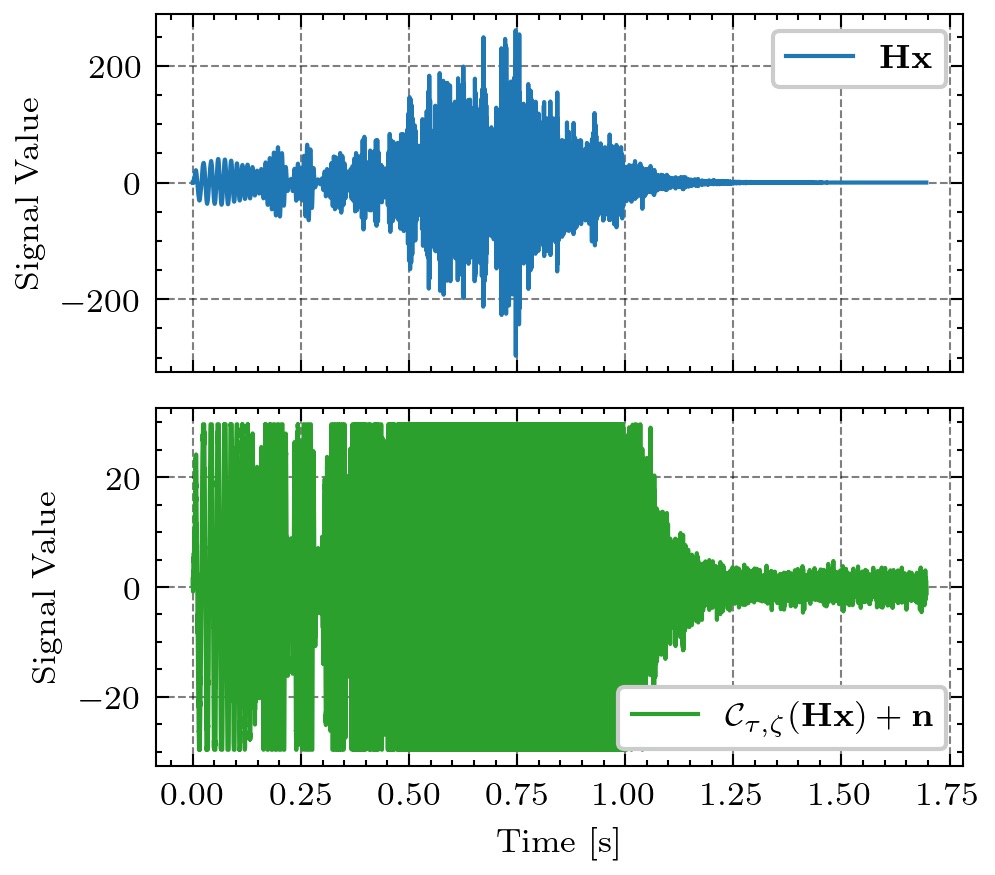}
\caption{Example of a measurement signal that is clipped. The convolved signal (top) is hard clipped and is contaminated with noise (bottom).}
\label{fig:clipping_example}
\end{figure}

A common nonlinear distortion in acoustic measurements is clipping, for instance, from limitations of the recording hardware and improperly adjusted recording levels. Although audio declipping has been studied extensively in the signal processing literature \cite{adler2011constrained, moliner2023solving}, its application to RIR estimation has received little attention to our knowledge. We consider the nonlinear inverse problem of clipped deconvolution, where the recorded reverberant signal is subject to clipping during measurement. Fig.~\ref{fig:clipping_example} shows an example of before and after clipping and also with added measurement noise. 

The measurement model is
\begin{align}
\mathbf{y} = \mathcal{C}_{\tau,\zeta}(\mathbf{H}\mathbf{x}) + \mathbf{n},
\qquad
\mathbf{n} \sim \mathcal{N}(\mathbf{0}, \sigma_n^2 \mathbf{I}),
\end{align}
where $\mathbf{x}\in\mathbb{R}^N$ is the unknown RIR, $\mathbf{h}\in\mathbb{R}^M$ is the excitation signal, $\mathbf{y}\in\mathbb{R}^{N+M-1}$ is the measured signal and $\mathcal{C}_{\tau,\zeta}(\cdot)$ denotes a smooth approximation of the hard clipping operator,
\begin{align}
\mathcal{C}_{\tau,\zeta}(x)
=
x
- \frac{1}{\zeta}\log\!\big(1 + e^{\zeta(x - \tau)}\big)
+ \frac{1}{\zeta}\log\!\big(1 + e^{\zeta(-x - \tau)}\big).
\end{align}
The parameter $\zeta>0$ determines the smoothness, in which larger value means better approximation to hard clipping, and the clipping threshold is set as
\begin{align}
\tau = c \max_i |y_i|,
\end{align}
where $c$ controls the clipping level. The smooth formulation is required to enable gradient-based optimization.

The measurement-aware refinement step is
\begin{align}
\begin{aligned}
\mu_t(\mathbf{x}_t,\mathbf{y})
&=
\arg\min_{\mathbf{x}}
\underbrace{
\frac{1}{2\sigma_n^2}
\big\|
\mathcal{C}_{\tau,\zeta}(\mathbf{H}\mathbf{x}) - \mathbf{y}
\big\|_2^2
}_{F_{\mathbf{y}}(\mathbf{x})}
\\ 
&+
\frac{1}{2\nu_t^2}
\|\mathbf{x}-\hat{\mathbf{x}}_1(\mathbf{x}_t)\|_2^2.
\label{eq:clipped_prox}
\end{aligned}
\end{align}

Due to the nonlinear forward model, the guided distribution
$\tilde{p}(\mathbf{X}_1 \mid \mathbf{X}_t=\mathbf{x}_t, \mathbf{Y}=\mathbf{y})$
is not a Gaussian. We therefore approximate it locally by a Gaussian distribution.
To this end, Eq.~\eqref{eq:clipped_prox} is solved using a Gauss–Newton (GN) method, in which the nonlinear forward model is linearized to solve quadratic subproblems.

At convergence, the objective admits a local quadratic approximation and gives
\begin{align}
\boldsymbol{\Sigma_t}^{-1}
=
\nu_t^{-2}\mathbf{I}
+
\sigma_n^{-2}\mathbf{J}^\top \mathbf{J},
\end{align}
where $\mathbf{J}$ denotes the Jacobian of the nonlinear operator
$\mathcal{C}_{\tau,\zeta}(\mathbf{X} \cdot)$ evaluated at the solution.
Both $\boldsymbol{\mu}_t$ and $\boldsymbol{\kappa}_t$ are again obtained by solving
linear systems using CG as in the $\ell_2$ and robust deconvolution cases.

\section{Results}\label{sec:experiments}

In this section, we evaluate the proposed inverse methods on a subset of 30 room impulse responses (RIRs) drawn from the MIT Acoustical Reverberation Scene Statistics Survey \cite{traer2016statistics}. 
The selected RIRs span RTs ranging from 0.2\,s to 2.3\,s, covering a wide variety of acoustic conditions. 
We use a sampling rate of $f_\mathrm{s}=8$\,kHz, which is sufficient for demonstrating the proposed approach. The denoiser is applied independently in 5 octave bands with center frequencies between 125\,Hz and 2\,kHz. 
To provide an intuitive illustration of the estimation result, we additionally present qualitative results on a representative RIR (``h054\_Kitchen\_3txts'', $\mathrm{RT}=0.5$\,s), which is used as a running example throughout the section.

\noindent\textbf{Evaluation Metric}: To quantify reconstruction performance, we employ the short-time normalized mean squared error (ST-NMSE). 
While the conventional NMSE measures the error over the entire signal, it is dominated by the high-energy early part of the RIR and therefore underrepresents errors in the late reverberation tail. Since accurate reconstruction of late decay is essential for all inverse problems considered in this work, we adopt a short-time formulation.

Let $\mathbf{x}_w, \hat{\mathbf{x}}_w \in \mathbb{R}^{N}$ denote the ground-truth and estimated RIR segments within the $w$-th analysis frame. The ST-NMSE is defined as
\begin{align}
\mathrm{ST\text{-}NMSE}_w
=
\frac{
\|\mathbf{x}_w - \hat{\mathbf{x}}_w\|_2^2
}{
\|\mathbf{x}_w\|_2^2
}.
\end{align}

We use a frame length of 20\,ms with a hop size of 10\,ms.  
The final performance metric, denoted $\mathrm{ST\text{-}NMSE}_{\mathrm{avg}}$, is obtained by averaging the ST-NMSE across all frames, such that early and late portions of the RIR contribute equally.

\noindent\textbf{Experimental Setup}: For all experiments, the number of flow steps is set to $T=1000$. 
We adopt a cosine time discretization so that for the $j$-th time step, 
\begin{align}
t = \frac{1 - \cos(\pi j / T)}{2},
\end{align}
which provides finer resolution towards the end.

The uncertainty $\gamma$ controls the trade-off between posterior sampling and deterministic reconstruction. While setting \(\gamma = 1\) corresponds to exact posterior sampling, previous work~\cite{pourya2025flower} has shown that \(\gamma = 0\) yields empirically the cleanest reconstructions. However, in our application of RIR inverse problems, this choice leads to over-denoising. We use \(\gamma = 0.9\), which gives the best result across all problems in this paper.


\subsection{Denoising}

We evaluate the denoising performance under three background noise conditions: 20~dB, 30~dB, and 40~dB SNR. The SNR is defined relative to the energy of the first 50~ms of the RIR, such that an SNR of 20~dB corresponds to a highly corrupted measurement in which the RIR is largely embedded in noise.

As a baseline for RIR denoising, we implement a \mbox{TruncShape} method, which is presented in similar forms on spatial RIR denoising \cite{masse2020robust, hold2022resynthesis}. The truncation is done by employing the classical Lundeby method~\cite{lundeby1995uncertainties} to estimate the RT and noise floor from the EDC. This determines a truncation point, which is where the noise floor is reached. Then, we resynthesize the response after the truncation point using shaped Gaussian noise with the an exponential decay determined by the estimated RT. We use the Lundeby estimator implementation provided in the \texttt{pyrato} toolbox.\footnote{\url{https://pyfar.org}} Owing to its conceptual simplicity and low computational cost, the TruncShape method is widely used in practice. However, it becomes unreliable at low SNRs, where even small errors in the estimated decay rate may lead to severe reconstruction artifacts or complete failure of the truncation point estimation. Moreover, ensuring that there is a seamless transition between the measured and synthesized signal segments is important. 

\begin{figure}[t!]
\centering
\includegraphics[width=\columnwidth]{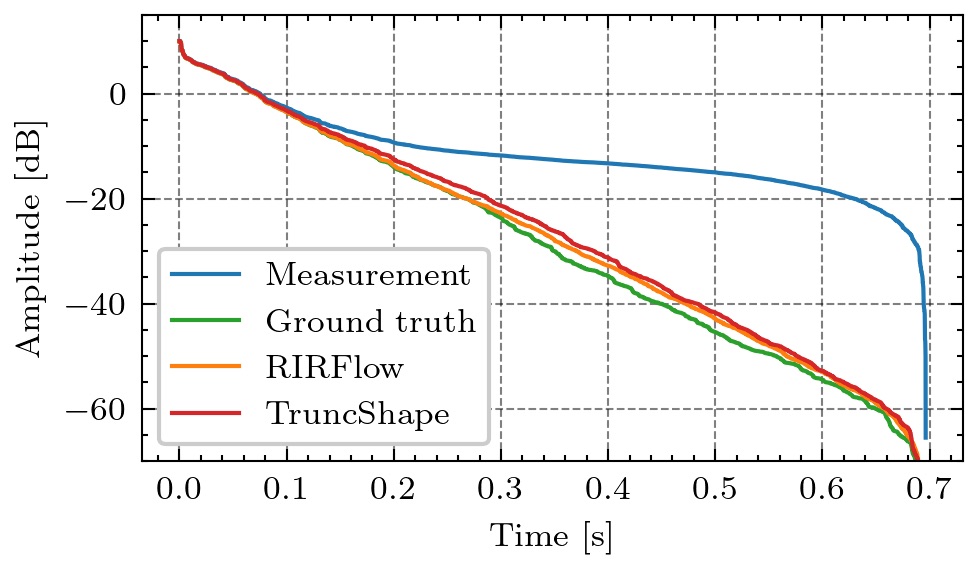}
\caption{EDCs for the denoising experiment on the RIR ``h054\_Kitchen\_3txts'' corrupted by Gaussian measurement noise with an SNR of 30~dB.}
\label{fig:denoising_example}
\end{figure}

\begin{figure}[t!]
\centering
\includegraphics[width=\columnwidth]{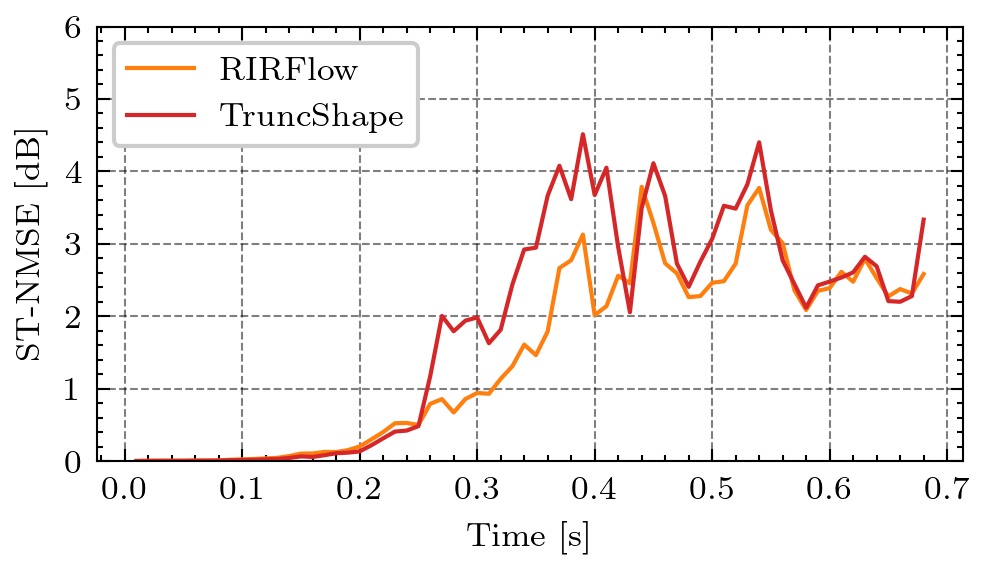}
\caption{ST-NMSE for the denoising task on the RIR ``h054\_Kitchen\_3txts'' corrupted by Gaussian measurement noise with an SNR of 30~dB, cf.~Fig.~\ref{fig:denoising_example}. Lower values correspond to more accurate energy reconstruction.}
\label{fig:denoising_ser_example}
\end{figure}

Figure~\ref{fig:denoising_example} shows an example of an RIR corrupted by 30~dB background noise. Both the TruncShape method and RIRFlow are able to suppress noise and recover the overall decay. However, the ST-NMSE curves in Fig.~\ref{fig:denoising_ser_example} show that the TruncShape method exhibits a higher error in the region between approximately 0.2~s and 0.4~s.

The observed error increase corresponds to the point where the measured signal is truncated and substituted with the synthesized decaying noise. This operation often introduces energy inconsistencies, which appear as artifacts in the reconstructed RIR. RIRFlow avoids this failure mode by sampling from a prior distribution over valid RIRs, ensuring that only physically plausible decay behavior is produced.

\begin{figure}[t!]
\centering
\includegraphics[width=\columnwidth]{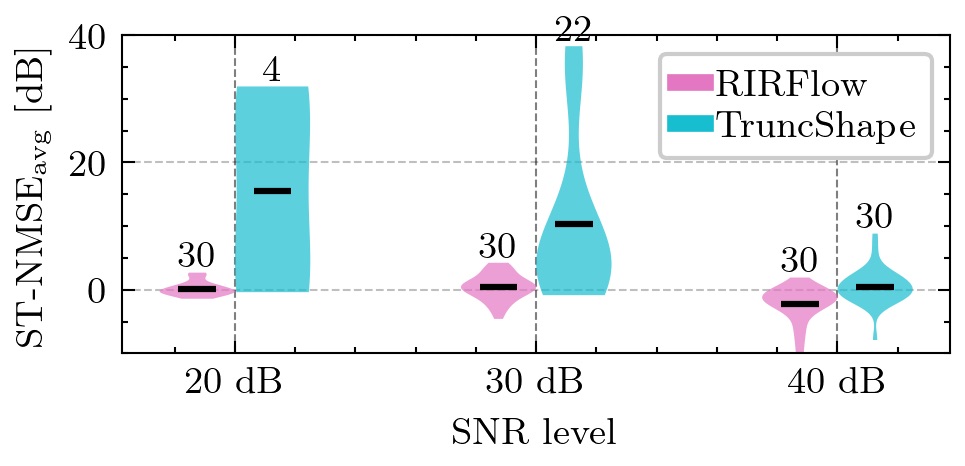}
\caption{Denoising performance on 30 RIRs under different measurement noise levels. The black line indicates the mean and the number above each violin plot shows the number of successful estimations out of 30 RIRs. The plot shows that the average errors of the proposed RIRFlow method are smaller than those of the Lundeby method. Lower ST-NMSE value indicates better performance.}
\label{fig:denoising_snr}
\end{figure}

We further evaluate denoising performance across 30 different RIRs and all three noise levels. The results are summarized in Fig.~\ref{fig:denoising_snr}, which also reports the number of successful estimates for each method. The TruncShape method fails to produce valid RT estimates under low-SNR conditions: only 4 and 22 successful estimates are obtained at 20~dB and 30~dB SNR, respectively, out of 30 RIRs. In contrast, RIRFlow successfully reconstructs all RIRs across all noise levels, demonstrating substantially improved robustness.

This behavior highlights a fundamental limitation of TruncShape in low-SNR conditions. When the RIR is heavily corrupted by noise, the linear regression underlying the TruncShape method becomes unstable, and even small perturbations can lead to severely biased or invalid RT estimates.
In contrast, RIRFlow's denoiser essentially performs an MMSE estimation of RT, as it infers by marginalizing over the full posterior distribution rather than relying on a single point estimate. Thus, MMSE formulation yields significantly more stable estimates under high uncertainty, making it inherently more robust in low-SNR conditions.

\begin{figure}[t!]
\centering
\includegraphics[width=\columnwidth]{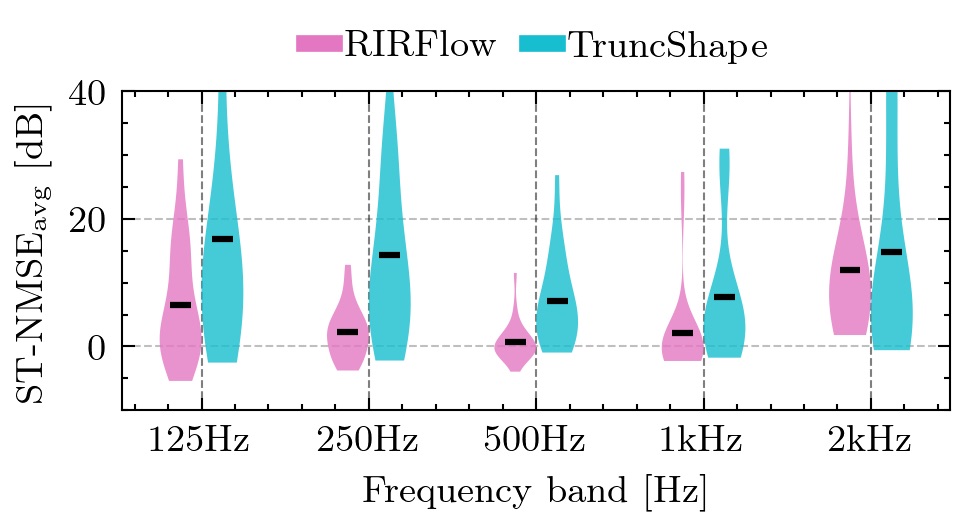}
\caption{Average ST-NMSE across octave bands for the denoising experiment corrupted by Gaussian measurement noise with an SNR of 30~dB. The black line indicates the mean.}
\label{fig:denoising_band}
\end{figure}

Finally, Fig.~\ref{fig:denoising_band} shows the ST-NMSE evaluated per frequency band at 30~dB SNR. While RIRFlow consistently outperforms the baseline, the results indicate that accurate band-wise reconstruction remains challenging and may introduce spectral coloration. 

\subsection{$\ell_2$ Deconvolution}

To simulate a sine-sweep measurement scenario, the observation signal is generated by convolving each RIR with a 1-s exponential sine sweep and subsequently adding Gaussian background noise. The noise level is set such that the resulting SNR is 30~dB, corresponding to a realistic noisy measurement condition. The noise level is determined from the specified SNR according to
\begin{align}
\sigma_n = \mathrm{RMS}(\mathbf{y}) \cdot 10^{-\mathrm{SNR}/20}.
\label{eq:snr}
\end{align}

\begin{figure}[t!]
\centering
\includegraphics[width=1.0\columnwidth]{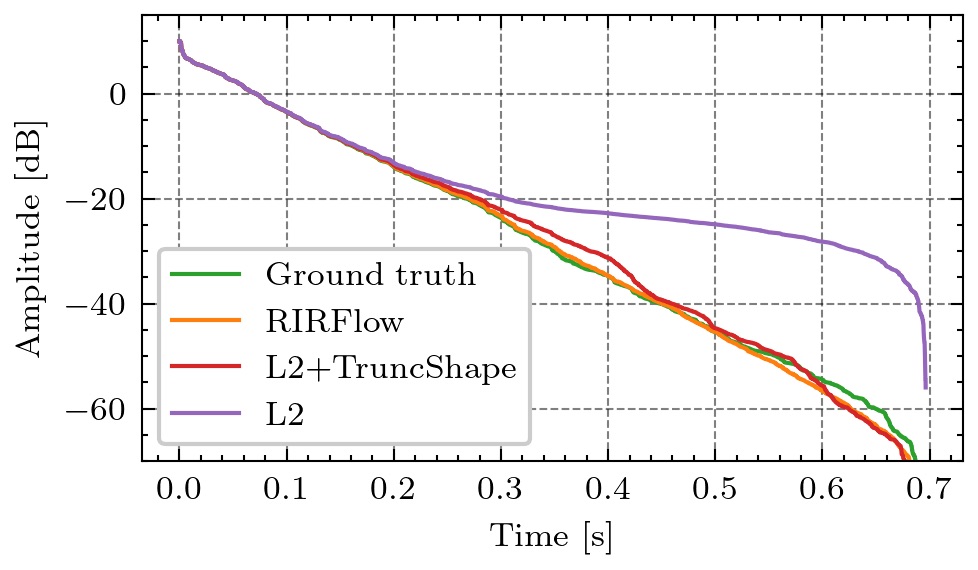}
\caption{EDCs for the $\ell_2$ deconvolution experiment on the RIR ``h054\_Kitchen\_3txts'' corrupted by Gaussian measurement noise with an SNR of 30~dB.\newline }
\label{fig:l2_deconv_example}
\end{figure}

The denoising result for the example RIR is shown in the EDC plot in Fig.\ref{fig:l2_deconv_example}. We compare RIRFlow against two baseline methods. The first baseline is L2, a classical $\ell_2$-norm deconvolution approach corresponding to a MLE estimator. This method solves only the data-fidelity term in Step~2 of RIRFlow, i.e., it minimizes $F_{\mathbf{y}}(\mathbf{x})$ in Eq.~\eqref{eq:l2_proximal}. The second baseline, denoted as L2+TruncShape, is L2 postprocessed with TruncShape to suppress late reverberation noise.
As expected, the plain L2 performs poorly, since it does not account for background noise. Applying the TruncShape method improves the results, which is consistent with its common use in practice to post-process by denoising the estimation. However, as in the previous denoising problem, there is a visible bump around the transition point to synthesized segments. RIRflow shows the smoothest and closest estimation result to the ground truth RIR.  

\begin{figure}[t!]
\centering
\includegraphics[width=\columnwidth]{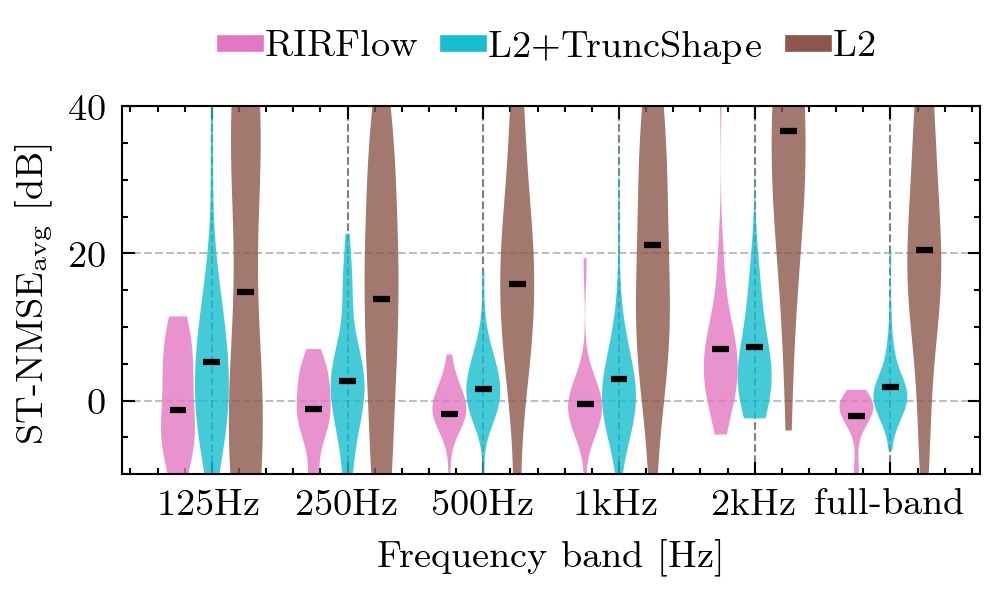}
\caption{Average ST-NMSE across octave bands for the $\ell_2$ deconvolution experiment corrupted by Gaussian measurement noise with an SNR of 30~dB.}
\label{fig:deconvL2_band}
\end{figure}

Fig.~\ref{fig:deconvL2_band} presents the per-band ST-NMSE error. RIRflow is able to estimate the energy of the ground truth RIR better than the other two methods. performance across all frequency bands. Again, while L2+TruncShape method also shows Unlike the two-stage procedure of deconvolution followed by denoising, the flow-based approach directly incorporates prior knowledge of clean RIR structure into the inversion process. As a result, the deconvolution process is aware of the background noise and thus the estimated RIR is inherently denoised. 

\subsection{Robust deconvolution}
Similar to the $\ell_2$ deconvolution experiment, we generate the observed signal by convolving a single exponential sine sweep with each test RIR. Additive noise is then introduced at an SNR of 30 dB for each of Gaussian and Laplacian noise, computed with respect to the RMS energy of the clean signal. Laplacian noise was generated with scale parameter $b = \sigma_\eta/\sqrt{2}$, where $\sigma_\eta$ is chosen to satisfy the desired SNR relative to the RMS energy of the clean signal in the same way as in Eq.~\eqref{eq:snr}. 

\begin{figure}[t!]
\centering
\includegraphics[width=1.0\columnwidth]{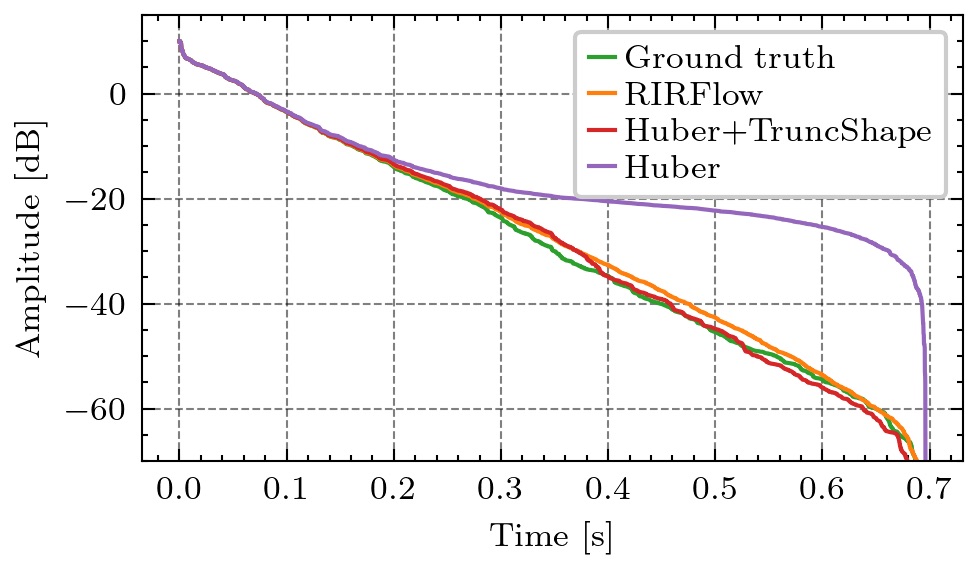}
\caption{EDCs for the robust deconvolution experiment on the RIR ``h054\_Kitchen\_3txts'' corrupted by Gaussian measurement noise with an SNR of 30~dB and Laplacian measurement noise with an SNR of 30~dB.}
\label{fig:l1_deconv_example}
\end{figure}

Figure~\ref{fig:l1_deconv_example} illustrates an example of the estimation result. Analogous to the baselines of $\ell_2$ case, the first baseline is an MLE estimator implemented with IRLS, referred to as Huber, and the second is Huber post-processed by TruncShape. In this example RIR, both FlowRIR and Huber+TruncShape method perform similarly well.  

\begin{figure}[t!]
\centering
\includegraphics[width=\columnwidth]{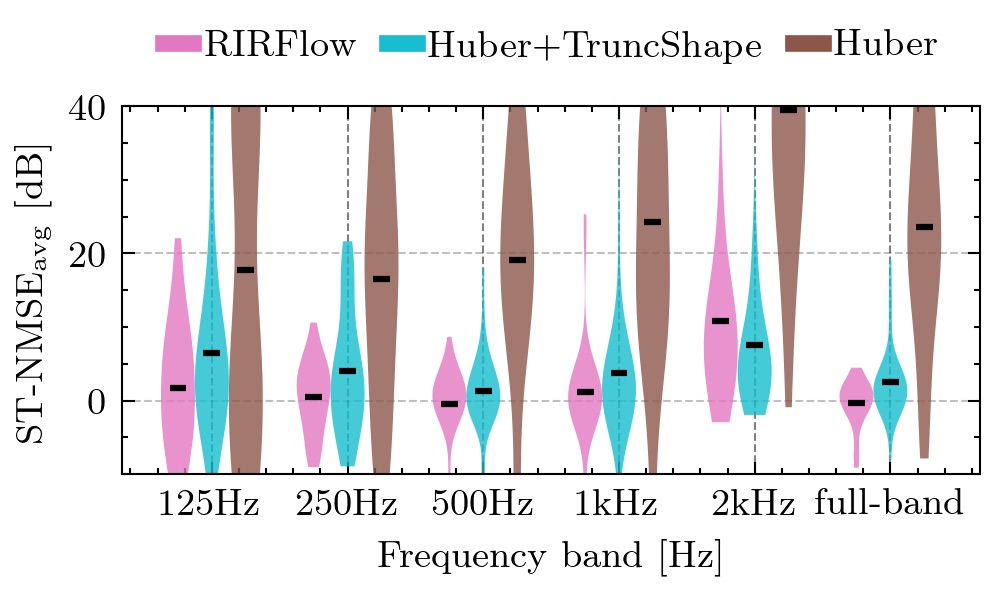}
\caption{Average ST-NMSE across octave bands for the robust deconvolution experiment corrupted by Gaussian measurement noise with an SNR of 30~dB and Laplacian measurement noise with an SNR of 30~dB.}
\label{fig:deconvL1_band}
\end{figure}

Result across different RIRs in Fig.~\ref{fig:deconvL1_band} shows a similar trend as in the $\ell_2$ deconvolution. RIRflow is able to estimate the true energy better and also has less variance in the error compared to the baseline methods. Although the true guided posterior $p(\mathbf{x}_1 \mid \mathbf{X}_t = \mathbf{x}_t, \mathbf{Y} = \mathbf{y})$ is not exactly Gaussian in this setting, the results indicate that the proposed Gaussian approximation captures the dominant structure of the posterior.

\subsection{Inpainting}
For the inpainting experiment, two temporal segments of the RIR are removed, spanning 0.1–0.2\,s and 0.5–1.0\,s, respectively. A negligible amount of additive noise (SNR = 100\,dB) is introduced to ensure numerical stability while keeping the unmasked regions effectively unchanged.

\begin{figure}[t!]
\centering
\includegraphics[width=\columnwidth]{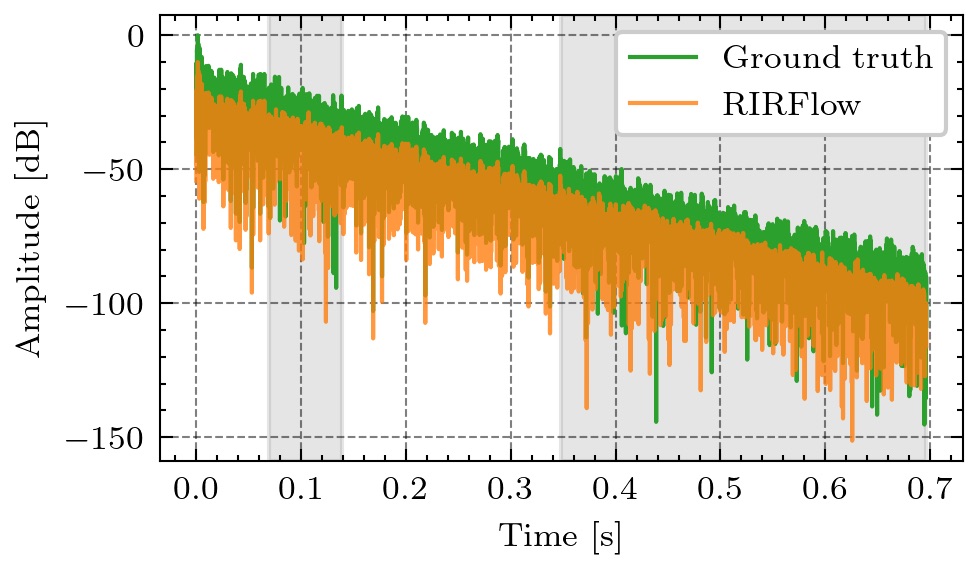}
\caption{Inpainting result on the cropped RIR (``h054\_Kitchen\_3txts'') from Fig.~\ref{fig:inpainting_example}. Gray shaded area is the cropped region. RIRFlow is able to reconstruct the energy of the ground truth RIR in this region.}
\label{fig:inpainting_result_example}
\end{figure}

Fig.~\ref{fig:inpainting_result_example} illustrates the inpainting result obtained with RIRFlow for the corrupted measurement shown in Fig.~\ref{fig:inpainting_example}. The proposed method successfully reconstructs the missing segments in a manner that is consistent with the clean portions of the RIR, producing a smooth and physically plausible segments.

The inpainting task particularly highlights the advantage of the proposed flow-based approach. In this setting, classical methods for estimating RT based on EDC are no longer feasible. By contrast, RIRFlow solves the inverse problem probabilistically through a sequence of controlled diffusion steps, allowing the reconstruction to be guided by the prior rather than by explicit decay fitting. Moreover, the MMSE formulation of the denoiser plays a key role in this scenario, as it provides a stable estimate of the RT even when large parts of the signal are unobserved.

\begin{figure}[t!]
\centering
\includegraphics[width=\columnwidth]{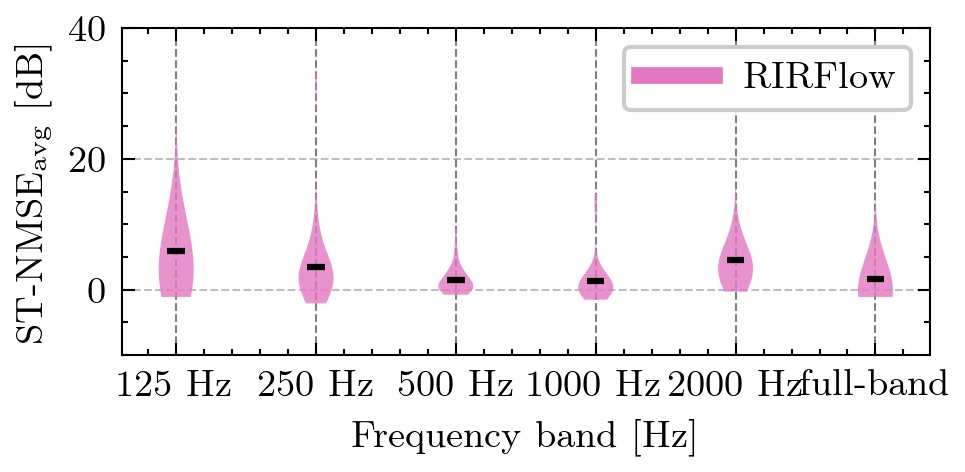}
\caption{Average ST-NMSE across octave bands for the RIR inpainting experiment.}
\label{fig:inpainting_band}
\end{figure}

To the best of our knowledge, no existing baseline method explicitly addresses the problem of RIR inpainting with arbitrarily missing temporal segments. For this reason, we report only the ST-NMSE results obtained using RIRFlow, shown in Fig.~\ref{fig:inpainting_band}.

\subsection{Declipping}

For the declipping experiment, the clipping threshold is set to $c = 0.1$ with a smoothing parameter $\zeta = 1000$. Additive Gaussian noise with an SNR of 30 dB is applied. During optimization, a smoothed approximation of the clipping operator is used to enable gradient-based updates, while the observed signal $\mathbf{y}$ is generated using hard clipping to reflect realistic recording conditions.

\begin{figure}[t!]
\centering
\includegraphics[width=\columnwidth]{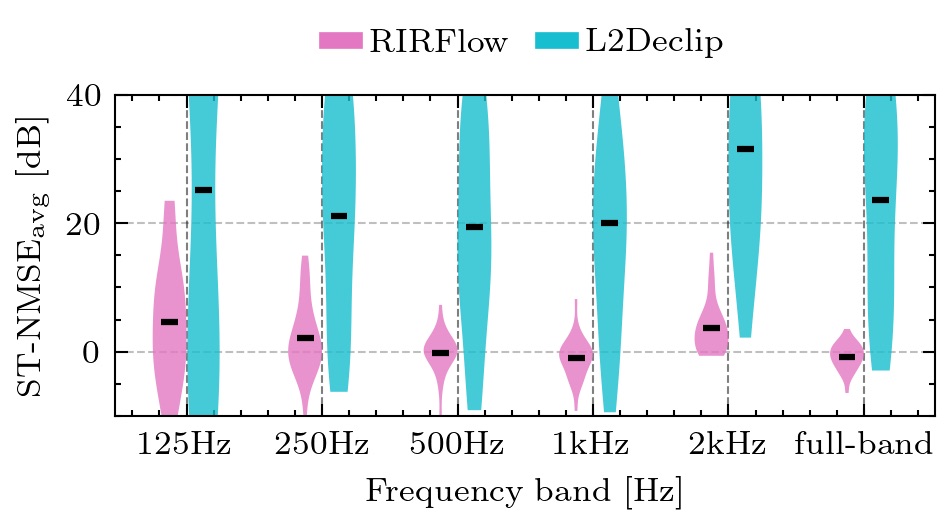}
\caption{Average ST-NMSE across octave bands for the declipping experiment corrupted by Gaussian measurement noise with an SNR of 30~dB.}
\label{fig:declipping_band}
\end{figure}

\begin{figure}[t!]
\centering
\includegraphics[width=\columnwidth]{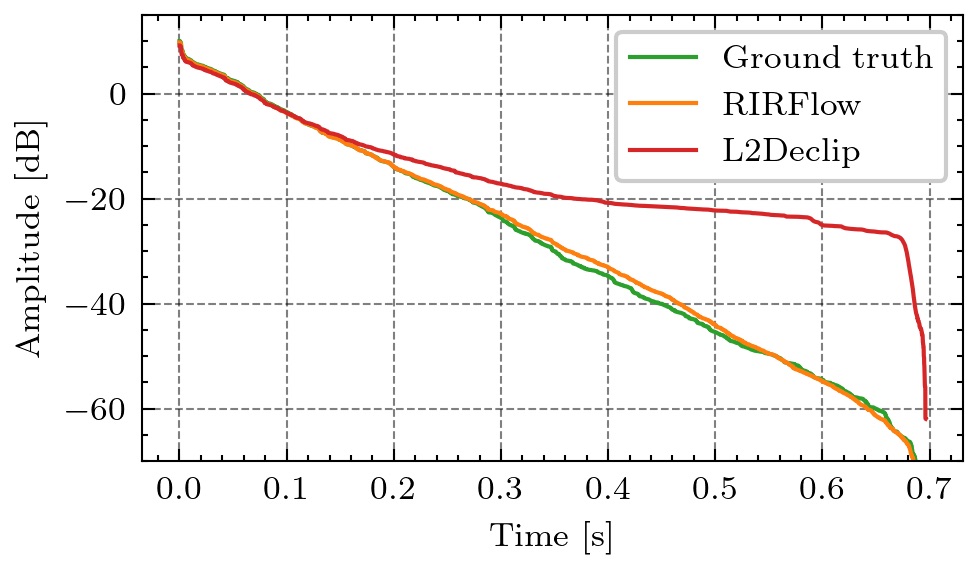}
\caption{EDCs for the declipping experiment on the RIR ``h054\_Kitchen\_3txts'' corrupted by Gaussian measurement noise with an SNR of 30~dB.}
\label{fig:declipping_example}
\end{figure}

As a baseline, we again consider an MLE estimator that minimizes only the data-fidelity term $F_{\mathbf{y}}(\mathbf{x})$ in Eq.~\eqref{eq:clipped_prox}, solved using a GN method. It is referred to as L2Declip. Fig.~\ref{fig:declipping_example} illustrates that RIRflow is able to suppress background noise and recover the overall energy decay of the RIR more faithfully than L2Declip. The quantitative comparison in Fig.~\ref{fig:declipping_band} further demonstrates that RIRflow achieves consistently lower ST-NMSE across frequency bands.

Declipping is a particularly challenging inverse problem in room acoustics, as the clipping operation introduces irreversible information loss and leads to a highly nonlinear and non-Gaussian observation model. As a result, classical optimization-based approaches often struggle to recover the underlying RIR, especially in the presence of noise.

RIRflow also exhibits limitations in this setting. Compared to other inverse problems considered earlier, declipping occasionally yields less stable reconstructions, with increased variability across different random initializations. This behavior can be attributed to the Gaussian approximation of the guided posterior distribution $p(\mathbf{x}_1 \mid \mathbf{X}_t = \mathbf{x}_t, \mathbf{Y} = \mathbf{y})$, which becomes less accurate for strongly nonlinear observation models such as clipping. Since the true posterior is non-Gaussian in this case, its structure cannot be fully captured by the first- and second-order moments alone, leading to increased variability in the sampled solutions. 

\section{Conclusion}\label{sec:conclusion}

This work introduced a unified Bayesian framework, RIRFlow, for solving RIR inverse problems using a flow-based solver with an analytic MMSE denoiser. By exploiting the statistical structure of RIRs, we derived a closed-form MMSE denoiser that serves as an explicit prior, enabling posterior sampling without any learned components. 
Compared to conventional approaches, the flow-based solver decomposes the challenging inverse problem into a sequence of simpler, tractable steps, leading to improved stability and robustness. The resulting framework unifies various inverse problems within a single framework, while remaining training-free and robust under challenging conditions. Experimental results demonstrate that combining analytic RIR modeling with flow-based inference yields consistent and stable estimations across a range of inverse problems, and achieves the best overall performance among the compared methods across all evaluated inverse problems.

Several directions remain for future work. The proposed analytic MMSE denoiser could be extended to incorporate more detailed frequency-dependent decay characteristics or to data-driven priors. In addition, practical measurement scenarios may need to measure or estimate the noise level $\sigma_n$. While this work demonstrates that the proposed framework remains effective for nonlinear and non-Gaussian inverse problems, further theoretical analysis of these extensions is a direction for future study. Finally, since the analytic MMSE denoiser is independent of the specific solver, it could be integrated with alternative flow- or diffusion-based solvers. Overall, this work demonstrates that a classical RIR model based on exponentially decaying variance can be successfully integrated into flow-based inverse solvers as a prior, enabling a unified and effective treatment of a broad class of RIR inverse problems.

\section{Author Declarations}
Conflict of Interest: The authors have no conflicts to disclose.
\begin{acknowledgments}
KYL and SJS were supported through the joint German Academic Exchange Service (DAAD) and Research Council of Finland (RCF) Project (57763119) "Immersive Augmented Acoustics (IAA)". Also, this work was supported by the HUCE infrastructure of the Aalto School of Electrical Engineering. 
\end{acknowledgments}









\bibliography{references}

\begin{thebibliography}{53}
\def\enquote#1{``#1,''}
\def\plainquote#1{``#1''}
\expandafter\ifx\csname natexlab\endcsname\relax\def\natexlab#1{#1}\fi
\providecommand{\dourl}[1]{\href{http://#1}{\nolinkurl{#1}}}
\providecommand{\bibinfo}[2]{#2}
\providecommand{\noopsort}[1]{}
\providecommand{\switchargs}[2]{#2#1}
  \def\eatspace #1{#1}

\bibitem[{Adler \emph{et~al.}(2011)Adler, Emiya, Jafari, Elad, Gribonval, and Plumbley}]{adler2011constrained}
\bibinfo{author}{Adler, A.}, \bibinfo{author}{Emiya, V.}, \bibinfo{author}{Jafari, M.~G.}, \bibinfo{author}{Elad, M.}, \bibinfo{author}{Gribonval, R.},  and \bibinfo{author}{Plumbley, M.~D.} (\textbf{\bibinfo{year}{2011}}). \enquote{\bibinfo{title}{A constrained matching pursuit approach to audio declipping}} in \emph{\bibinfo{booktitle}{Proc. IEEE Int. Conf. Acoust. Speech Signal Process. (ICASSP)}}, \bibinfo{organization}{IEEE}, pp. \bibinfo{pages}{329--332}.

\bibitem[{Antonello \emph{et~al.}(2017)Antonello, De~Sena, Moonen, Naylor, and Van~Waterschoot}]{antonello2017room}
\bibinfo{author}{Antonello, N.}, \bibinfo{author}{De~Sena, E.}, \bibinfo{author}{Moonen, M.}, \bibinfo{author}{Naylor, P.~A.},  and \bibinfo{author}{Van~Waterschoot, T.} (\textbf{\bibinfo{year}{2017}}). \enquote{\bibinfo{title}{Room impulse response interpolation using a sparse spatio-temporal representation of the sound field}} \bibinfo{journal}{IEEE Trans. Audio Speech Lang. Process.} \textbf{25}(10), \bibinfo{pages}{1929--1941}.

\bibitem[{Blau and Michaeli(2018)}]{blau2018perception}
\bibinfo{author}{Blau, Y.},  and \bibinfo{author}{Michaeli, T.} (\textbf{\bibinfo{year}{2018}}). \enquote{\bibinfo{title}{The perception-distortion tradeoff}} in \emph{\bibinfo{booktitle}{Proceedings of the IEEE conference on computer vision and pattern recognition}}, pp. \bibinfo{pages}{6228--6237}.

\bibitem[{Boyd and Vandenberghe(2004)}]{boyd2004convex}
\bibinfo{author}{Boyd, S.},  and \bibinfo{author}{Vandenberghe, L.} (\textbf{\bibinfo{year}{2004}}). \emph{\bibinfo{title}{Convex optimization}} (\bibinfo{publisher}{Cambridge university press}).

\bibitem[{Cabrera \emph{et~al.}(2011)Cabrera, Lee, Yadav, and Martens}]{cabrera2011decay}
\bibinfo{author}{Cabrera, D.}, \bibinfo{author}{Lee, D.}, \bibinfo{author}{Yadav, M.},  and \bibinfo{author}{Martens, W.~L.} (\textbf{\bibinfo{year}{2011}}). \enquote{\bibinfo{title}{Decay envelope manipulation of room impulse responses: Techniques for auralization and sonification}} in \emph{\bibinfo{booktitle}{Proceedings of ACOUSTICS}}, p.~\bibinfo{pages}{5}.

\bibitem[{Cang \emph{et~al.}(2022)Cang, Sheng, Jakobsson, and Yang}]{cang2022robust}
\bibinfo{author}{Cang, S.}, \bibinfo{author}{Sheng, X.}, \bibinfo{author}{Jakobsson, A.},  and \bibinfo{author}{Yang, H.} (\textbf{\bibinfo{year}{2022}}). \enquote{\bibinfo{title}{Robust deconvolution of underwater acoustic channels corrupted by impulsive noise}} in \emph{\bibinfo{booktitle}{2022 5th International Conference on Information Communication and Signal Processing (ICICSP)}}, \bibinfo{organization}{IEEE}, pp. \bibinfo{pages}{571--576}.

\bibitem[{Caviedes-Nozal and Fernandez-Grande(2023)}]{caviedes2023spatio}
\bibinfo{author}{Caviedes-Nozal, D.},  and \bibinfo{author}{Fernandez-Grande, E.} (\textbf{\bibinfo{year}{2023}}). \enquote{\bibinfo{title}{Spatio-temporal {Bayesian} regression for room impulse response reconstruction with spherical waves}} \bibinfo{journal}{IEEE Trans. Audio Speech Lang. Process.} \textbf{31}, \bibinfo{pages}{3263--3277}.

\bibitem[{Chung \emph{et~al.}(2022)Chung, Kim, Mccann, Klasky, and Ye}]{chung2022diffusion}
\bibinfo{author}{Chung, H.}, \bibinfo{author}{Kim, J.}, \bibinfo{author}{Mccann, M.~T.}, \bibinfo{author}{Klasky, M.~L.},  and \bibinfo{author}{Ye, J.~C.} (\textbf{\bibinfo{year}{2022}}). \enquote{\bibinfo{title}{Diffusion posterior sampling for general noisy inverse problems}} \bibinfo{journal}{arXiv preprint arXiv:2209.14687} .

\bibitem[{Crocco and Del~Bue(2015)}]{crocco2015room}
\bibinfo{author}{Crocco, M.},  and \bibinfo{author}{Del~Bue, A.} (\textbf{\bibinfo{year}{2015}}). \enquote{\bibinfo{title}{Room impulse response estimation by iterative weighted l 1-norm}} in \emph{\bibinfo{booktitle}{Proc. European Signal Process. Conf. (EUSIPCO)}}, pp. \bibinfo{pages}{1895--1899}.

\bibitem[{Efron(2011)}]{efron2011tweedie}
\bibinfo{author}{Efron, B.} (\textbf{\bibinfo{year}{2011}}). \enquote{\bibinfo{title}{Tweedie’s formula and selection bias}} \bibinfo{journal}{Journal of the American Statistical Association} \textbf{106}(496), \bibinfo{pages}{1602--1614}.

\bibitem[{Farina(2000)}]{farina_simultaneous_2000}
\bibinfo{author}{Farina, A.} (\textbf{\bibinfo{year}{2000}}). \enquote{\bibinfo{title}{Simultaneous measurement of impulse response and distortion with a swept-sine technique}} in \emph{\bibinfo{booktitle}{Proc. 108th {AES} {Convention}}}, \bibinfo{address}{Paris, France}.

\bibitem[{Gagneux \emph{et~al.}(2025)Gagneux, Martin, Gribonval, and Massias}]{gagneux2025generation}
\bibinfo{author}{Gagneux, A.}, \bibinfo{author}{Martin, S.}, \bibinfo{author}{Gribonval, R.},  and \bibinfo{author}{Massias, M.} (\textbf{\bibinfo{year}{2025}}). \enquote{\bibinfo{title}{The generation phases of flow matching: a denoising perspective}} \bibinfo{journal}{arXiv preprint arXiv:2510.24830} .

\bibitem[{Goodwin and Jot(2006)}]{goodwin2006analysis}
\bibinfo{author}{Goodwin, M.~M.},  and \bibinfo{author}{Jot, J.-M.} (\textbf{\bibinfo{year}{2006}}). \enquote{\bibinfo{title}{Analysis and synthesis for universal spatial audio coding}} in \emph{\bibinfo{booktitle}{Audio Engineering Society Convention 121}}, \bibinfo{organization}{Audio Engineering Society}.

\bibitem[{Hold \emph{et~al.}(2022)Hold, McKenzie, G{\"o}tz, Schlecht, and Pulkki}]{hold2022resynthesis}
\bibinfo{author}{Hold, C.}, \bibinfo{author}{McKenzie, T.}, \bibinfo{author}{G{\"o}tz, G.}, \bibinfo{author}{Schlecht, S.},  and \bibinfo{author}{Pulkki, V.} (\textbf{\bibinfo{year}{2022}}). \enquote{\bibinfo{title}{Resynthesis of spatial room impulse response tails with anisotropic multi-slope decays}} \bibinfo{journal}{journal of the audio engineering society} \textbf{70}(6), \bibinfo{pages}{526--538}.

\bibitem[{Holderrieth and Erives(2025)}]{holderrieth2025introduction}
\bibinfo{author}{Holderrieth, P.},  and \bibinfo{author}{Erives, E.} (\textbf{\bibinfo{year}{2025}}). \enquote{\bibinfo{title}{An introduction to flow matching and diffusion models}} \bibinfo{journal}{arXiv preprint arXiv:2506.02070} .

\bibitem[{J{\"a}lmby \emph{et~al.}(2023)J{\"a}lmby, Elvander, and Van~Waterschoot}]{jalmby2023low}
\bibinfo{author}{J{\"a}lmby, M.}, \bibinfo{author}{Elvander, F.},  and \bibinfo{author}{Van~Waterschoot, T.} (\textbf{\bibinfo{year}{2023}}). \enquote{\bibinfo{title}{Low-rank room impulse response estimation}} \bibinfo{journal}{IEEE Trans. Audio Speech Lang. Process.} \textbf{31}, \bibinfo{pages}{957--969}.

\bibitem[{Karakonstantis \emph{et~al.}(2024)Karakonstantis, Caviedes-Nozal, Richard, and Fernandez-Grande}]{karakonstantis2024room}
\bibinfo{author}{Karakonstantis, X.}, \bibinfo{author}{Caviedes-Nozal, D.}, \bibinfo{author}{Richard, A.},  and \bibinfo{author}{Fernandez-Grande, E.} (\textbf{\bibinfo{year}{2024}}). \enquote{\bibinfo{title}{Room impulse response reconstruction with physics-informed deep learning}} \bibinfo{journal}{The Journal of the Acoustical Society of America} \textbf{155}(2), \bibinfo{pages}{1048--1059}.

\bibitem[{Kawar \emph{et~al.}(2022)Kawar, Song, Ermon, and Elad}]{kawar2022jpeg}
\bibinfo{author}{Kawar, B.}, \bibinfo{author}{Song, J.}, \bibinfo{author}{Ermon, S.},  and \bibinfo{author}{Elad, M.} (\textbf{\bibinfo{year}{2022}}). \enquote{\bibinfo{title}{{JPEG} artifact correction using denoising diffusion restoration models}} \bibinfo{journal}{arXiv preprint arXiv:2209.11888} .

\bibitem[{Kreutz \emph{et~al.}(2013)Kreutz, Raue, Kaschek, and Timmer}]{kreutz2013profile}
\bibinfo{author}{Kreutz, C.}, \bibinfo{author}{Raue, A.}, \bibinfo{author}{Kaschek, D.},  and \bibinfo{author}{Timmer, J.} (\textbf{\bibinfo{year}{2013}}). \enquote{\bibinfo{title}{Profile likelihood in systems biology}} \bibinfo{journal}{The FEBS journal} \textbf{280}(11), \bibinfo{pages}{2564--2571}.

\bibitem[{Kuttruff(2009)}]{Kuttruff2009RA}
\bibinfo{author}{Kuttruff, H.} (\textbf{\bibinfo{year}{2009}}). CRC Press \emph{\bibinfo{title}{{Room Acoustics}}}, \bibinfo{edition}{5th} ed. (\bibinfo{publisher}{CRC Press}).

\bibitem[{Lee \emph{et~al.}(2026)Lee, Meyer-Kahlen, Prawda, Välimäki, and Schlecht}]{lee2026anyrir}
\bibinfo{author}{Lee, K.~Y.}, \bibinfo{author}{Meyer-Kahlen, N.}, \bibinfo{author}{Prawda, K.}, \bibinfo{author}{Välimäki, V.},  and \bibinfo{author}{Schlecht, S.~J.} (\textbf{\bibinfo{year}{2026}}). \enquote{\bibinfo{title}{{AnyRIR}: Robust non-intrusive room impulse response estimation in the wild}} in \emph{\bibinfo{booktitle}{Proc. IEEE Int. Conf. Acoust. Speech Signal Process. (ICASSP)}}, \bibinfo{organization}{IEEE}, pp. \bibinfo{pages}{1--5}.

\bibitem[{Lemercier \emph{et~al.}(2025)Lemercier, Moliner, Welker, V{\"a}lim{\"a}ki, and Gerkmann}]{lemercier2025unsupervised}
\bibinfo{author}{Lemercier, J.-M.}, \bibinfo{author}{Moliner, E.}, \bibinfo{author}{Welker, S.}, \bibinfo{author}{V{\"a}lim{\"a}ki, V.},  and \bibinfo{author}{Gerkmann, T.} (\textbf{\bibinfo{year}{2025}}). \enquote{\bibinfo{title}{Unsupervised blind joint dereverberation and room acoustics estimation with diffusion models}} \bibinfo{journal}{IEEE Trans. Audio Speech Lang. Process.} \textbf{33}, \bibinfo{pages}{2244--2258}.

\bibitem[{Lin and Lee(2006)}]{lin2006bayesian}
\bibinfo{author}{Lin, Y.},  and \bibinfo{author}{Lee, D.~D.} (\textbf{\bibinfo{year}{2006}}). \enquote{\bibinfo{title}{Bayesian regularization and nonnegative deconvolution for room impulse response estimation}} \bibinfo{journal}{IEEE Transactions on Signal Processing} \textbf{54}(3), \bibinfo{pages}{839--847}.

\bibitem[{Lipman \emph{et~al.}(2022)Lipman, Chen, Ben-Hamu, Nickel, and Le}]{lipman2022flow}
\bibinfo{author}{Lipman, Y.}, \bibinfo{author}{Chen, R.~T.}, \bibinfo{author}{Ben-Hamu, H.}, \bibinfo{author}{Nickel, M.},  and \bibinfo{author}{Le, M.} (\textbf{\bibinfo{year}{2022}}). \enquote{\bibinfo{title}{Flow matching for generative modeling}} \bibinfo{journal}{arXiv preprint arXiv:2210.02747} .

\bibitem[{Liu \emph{et~al.}(2022)Liu, Gong, and Liu}]{liu2022flow}
\bibinfo{author}{Liu, X.}, \bibinfo{author}{Gong, C.},  and \bibinfo{author}{Liu, Q.} (\textbf{\bibinfo{year}{2022}}). \enquote{\bibinfo{title}{Flow straight and fast: Learning to generate and transfer data with rectified flow}} \bibinfo{journal}{arXiv preprint arXiv:2209.03003} .

\bibitem[{Lluis \emph{et~al.}(2020)Lluis, Martinez-Nuevo, Bo~M{\o}ller, and Ewan~Shepstone}]{lluis2020sound}
\bibinfo{author}{Lluis, F.}, \bibinfo{author}{Martinez-Nuevo, P.}, \bibinfo{author}{Bo~M{\o}ller, M.},  and \bibinfo{author}{Ewan~Shepstone, S.} (\textbf{\bibinfo{year}{2020}}). \enquote{\bibinfo{title}{Sound field reconstruction in rooms: Inpainting meets super-resolution}} \bibinfo{journal}{The Journal of the Acoustical Society of America} \textbf{148}(2), \bibinfo{pages}{649--659}.

\bibitem[{Lundeby \emph{et~al.}(1995)Lundeby, Vigran, Bietz, and Vorl{\"a}nder}]{lundeby1995uncertainties}
\bibinfo{author}{Lundeby, A.}, \bibinfo{author}{Vigran, T.~E.}, \bibinfo{author}{Bietz, H.},  and \bibinfo{author}{Vorl{\"a}nder, M.} (\textbf{\bibinfo{year}{1995}}). \enquote{\bibinfo{title}{Uncertainties of measurements in room acoustics}} \bibinfo{journal}{Acta Acustica united with Acustica} \textbf{81}(4), \bibinfo{pages}{344--355}.

\bibitem[{Martin \emph{et~al.}(2024)Martin, Gagneux, Hagemann, and Steidl}]{martin2024pnp}
\bibinfo{author}{Martin, S.}, \bibinfo{author}{Gagneux, A.}, \bibinfo{author}{Hagemann, P.},  and \bibinfo{author}{Steidl, G.} (\textbf{\bibinfo{year}{2024}}). \enquote{\bibinfo{title}{Pnp-flow: Plug-and-play image restoration with flow matching}} \bibinfo{journal}{arXiv preprint arXiv:2410.02423} .

\bibitem[{Mass{\'e} \emph{et~al.}(2020)Mass{\'e}, Carpentier, Warusfel, and Noisternig}]{masse2020robust}
\bibinfo{author}{Mass{\'e}, P.}, \bibinfo{author}{Carpentier, T.}, \bibinfo{author}{Warusfel, O.},  and \bibinfo{author}{Noisternig, M.} (\textbf{\bibinfo{year}{2020}}). \enquote{\bibinfo{title}{A robust denoising process for spatial room impulse responses with diffuse reverberation tails}} \bibinfo{journal}{The Journal of the Acoustical Society of America} \textbf{147}(4), \bibinfo{pages}{2250--2260}.

\bibitem[{Moliner \emph{et~al.}(2023)Moliner, Lehtinen, and V{\"a}lim{\"a}ki}]{moliner2023solving}
\bibinfo{author}{Moliner, E.}, \bibinfo{author}{Lehtinen, J.},  and \bibinfo{author}{V{\"a}lim{\"a}ki, V.} (\textbf{\bibinfo{year}{2023}}). \enquote{\bibinfo{title}{Solving audio inverse problems with a diffusion model}} in \emph{\bibinfo{booktitle}{Proc. IEEE Int. Conf. Acoust. Speech Signal Process. (ICASSP)}}, \bibinfo{organization}{IEEE}, pp. \bibinfo{pages}{1--5}.

\bibitem[{Müller and Massarani(2001)}]{muller_transfer-function_2001}
\bibinfo{author}{Müller, S.},  and \bibinfo{author}{Massarani, P.} (\textbf{\bibinfo{year}{2001}}). \enquote{\bibinfo{title}{Transfer-function measurement with sweeps}} \bibinfo{journal}{J. Audio Eng. Soc.} \textbf{49}(6), \bibinfo{pages}{29}.

\bibitem[{Papadakis and Stavroulakis(2020)}]{papadakis_handclap_2020}
\bibinfo{author}{Papadakis, N.~M.},  and \bibinfo{author}{Stavroulakis, G.~E.} (\textbf{\bibinfo{year}{2020}}). \enquote{\bibinfo{title}{Handclap for {Acoustic} {Measurements}: {Optimal} {Application} and {Limitations}}} \bibinfo{journal}{Acoustics} \textbf{2}(2), \bibinfo{pages}{224--245}, \dourl{https://www.mdpi.com/2624-599X/2/2/15}, \dodoi{10.3390/acoustics2020015}.

\bibitem[{Parikh \emph{et~al.}(2014)Parikh, Boyd \emph{et~al.}}]{parikh2014proximal}
\bibinfo{author}{Parikh, N.}, \bibinfo{author}{Boyd, S.} \emph{et~al.} (\textbf{\bibinfo{year}{2014}}). \enquote{\bibinfo{title}{Proximal algorithms}} \bibinfo{journal}{Foundations and trends{\textregistered} in Optimization} \textbf{1}(3), \bibinfo{pages}{127--239}.

\bibitem[{P{\"a}tynen \emph{et~al.}(2011)P{\"a}tynen, Katz, and Lokki}]{patynen2011investigations}
\bibinfo{author}{P{\"a}tynen, J.}, \bibinfo{author}{Katz, B.~F.},  and \bibinfo{author}{Lokki, T.} (\textbf{\bibinfo{year}{2011}}). \enquote{\bibinfo{title}{Investigations on the balloon as an impulse source}} \bibinfo{journal}{The Journal of the Acoustical Society of America} \textbf{129}(1), \bibinfo{pages}{EL27--EL33}.

\bibitem[{Pezzoli \emph{et~al.}(2022)Pezzoli, Perini, Bernardini, Borra, Antonacci, and Sarti}]{pezzoli2022deep}
\bibinfo{author}{Pezzoli, M.}, \bibinfo{author}{Perini, D.}, \bibinfo{author}{Bernardini, A.}, \bibinfo{author}{Borra, F.}, \bibinfo{author}{Antonacci, F.},  and \bibinfo{author}{Sarti, A.} (\textbf{\bibinfo{year}{2022}}). \enquote{\bibinfo{title}{Deep prior approach for room impulse response reconstruction}} \bibinfo{journal}{Sensors} \textbf{22}(7), \bibinfo{pages}{2710}.

\bibitem[{Pokle \emph{et~al.}(2023)Pokle, Muckley, Chen, and Karrer}]{pokle2023training}
\bibinfo{author}{Pokle, A.}, \bibinfo{author}{Muckley, M.~J.}, \bibinfo{author}{Chen, R.~T.},  and \bibinfo{author}{Karrer, B.} (\textbf{\bibinfo{year}{2023}}). \enquote{\bibinfo{title}{Training-free linear image inverses via flows}} \bibinfo{journal}{arXiv preprint arXiv:2310.04432} .

\bibitem[{Pourya \emph{et~al.}(2025)Pourya, Rawas, and Unser}]{pourya2025flower}
\bibinfo{author}{Pourya, M.}, \bibinfo{author}{Rawas, B.~E.},  and \bibinfo{author}{Unser, M.} (\textbf{\bibinfo{year}{2025}}). \enquote{\bibinfo{title}{Flower: A flow-matching solver for inverse problems}} \bibinfo{journal}{arXiv preprint arXiv:2509.26287} .

\bibitem[{Prawda \emph{et~al.}(2025)Prawda, Meyer-Kahlen, and Schlecht}]{prawda2025cropping}
\bibinfo{author}{Prawda, K.}, \bibinfo{author}{Meyer-Kahlen, N.},  and \bibinfo{author}{Schlecht, S.~J.} (\textbf{\bibinfo{year}{2025}}). \enquote{\bibinfo{title}{Cropping room impulse responses using unimodal regression of their covariance}} \bibinfo{journal}{JASA Express Lett.} \textbf{5}(8).

\bibitem[{Richard \emph{et~al.}(2022)Richard, Dodds, and Ithapu}]{richard2022deep}
\bibinfo{author}{Richard, A.}, \bibinfo{author}{Dodds, P.},  and \bibinfo{author}{Ithapu, V.~K.} (\textbf{\bibinfo{year}{2022}}). \enquote{\bibinfo{title}{Deep impulse responses: Estimating and parameterizing filters with deep networks}} in \emph{\bibinfo{booktitle}{Proc. IEEE Int. Conf. Acoust. Speech Signal Process. (ICASSP)}}, \bibinfo{organization}{IEEE}, pp. \bibinfo{pages}{3209--3213}.

\bibitem[{Romano \emph{et~al.}(2017)Romano, Elad, and Milanfar}]{romano2017little}
\bibinfo{author}{Romano, Y.}, \bibinfo{author}{Elad, M.},  and \bibinfo{author}{Milanfar, P.} (\textbf{\bibinfo{year}{2017}}). \enquote{\bibinfo{title}{The little engine that could: Regularization by denoising (red)}} \bibinfo{journal}{SIAM journal on imaging sciences} \textbf{10}(4), \bibinfo{pages}{1804--1844}.

\bibitem[{Sohl-Dickstein \emph{et~al.}(2015)Sohl-Dickstein, Weiss, Maheswaranathan, and Ganguli}]{sohl2015deep}
\bibinfo{author}{Sohl-Dickstein, J.}, \bibinfo{author}{Weiss, E.}, \bibinfo{author}{Maheswaranathan, N.},  and \bibinfo{author}{Ganguli, S.} (\textbf{\bibinfo{year}{2015}}). \enquote{\bibinfo{title}{Deep unsupervised learning using nonequilibrium thermodynamics}} in \emph{\bibinfo{booktitle}{International conference on machine learning}}, \bibinfo{organization}{pmlr}, pp. \bibinfo{pages}{2256--2265}.

\bibitem[{Song \emph{et~al.}(2023)Song, Vahdat, Mardani, and Kautz}]{song2023pseudoinverse}
\bibinfo{author}{Song, J.}, \bibinfo{author}{Vahdat, A.}, \bibinfo{author}{Mardani, M.},  and \bibinfo{author}{Kautz, J.} (\textbf{\bibinfo{year}{2023}}). \enquote{\bibinfo{title}{Pseudoinverse-guided diffusion models for inverse problems}} in \emph{\bibinfo{booktitle}{International Conference on Learning Representations}}.

\bibitem[{Song and Ermon(2019)}]{song2019generative}
\bibinfo{author}{Song, Y.},  and \bibinfo{author}{Ermon, S.} (\textbf{\bibinfo{year}{2019}}). \enquote{\bibinfo{title}{Generative modeling by estimating gradients of the data distribution}} \bibinfo{journal}{Advances in neural information processing systems} \textbf{32}.

\bibitem[{Song \emph{et~al.}(2020)Song, Sohl-Dickstein, Kingma, Kumar, Ermon, and Poole}]{song2020score}
\bibinfo{author}{Song, Y.}, \bibinfo{author}{Sohl-Dickstein, J.}, \bibinfo{author}{Kingma, D.~P.}, \bibinfo{author}{Kumar, A.}, \bibinfo{author}{Ermon, S.},  and \bibinfo{author}{Poole, B.} (\textbf{\bibinfo{year}{2020}}). \enquote{\bibinfo{title}{Score-based generative modeling through stochastic differential equations}} \bibinfo{journal}{arXiv preprint arXiv:2011.13456} .

\bibitem[{Sundstr{\"o}m \emph{et~al.}(2024)Sundstr{\"o}m, Elvander, and Jakobsson}]{sundstrom2024estimation}
\bibinfo{author}{Sundstr{\"o}m, D.}, \bibinfo{author}{Elvander, F.},  and \bibinfo{author}{Jakobsson, A.} (\textbf{\bibinfo{year}{2024}}). \enquote{\bibinfo{title}{Estimation of impulse responses for a moving source using optimal transport regularization}} in \emph{\bibinfo{booktitle}{Proc. IEEE Int. Conf. Acoust. Speech Signal Process. (ICASSP)}}, \bibinfo{organization}{IEEE}, pp. \bibinfo{pages}{921--925}.

\bibitem[{Traer and McDermott(2016)}]{traer2016statistics}
\bibinfo{author}{Traer, J.},  and \bibinfo{author}{McDermott, J.~H.} (\textbf{\bibinfo{year}{2016}}). \enquote{\bibinfo{title}{Statistics of natural reverberation enable perceptual separation of sound and space}} \bibinfo{journal}{Proceedings of the National Academy of Sciences} \textbf{113}(48), \bibinfo{pages}{E7856--E7865}.

\bibitem[{van Waterschoot(2025)}]{van2025deep}
\bibinfo{author}{van Waterschoot, T.} (\textbf{\bibinfo{year}{2025}}). \enquote{\bibinfo{title}{Deep, data-driven modeling of room acoustics: literature review and research perspectives}} \bibinfo{journal}{arXiv preprint arXiv:2504.16289} .

\bibitem[{van Waterschoot \emph{et~al.}(2008)van Waterschoot, Rombouts, and Moonen}]{van2008optimally}
\bibinfo{author}{van Waterschoot, T.}, \bibinfo{author}{Rombouts, G.},  and \bibinfo{author}{Moonen, M.} (\textbf{\bibinfo{year}{2008}}). \enquote{\bibinfo{title}{Optimally regularized adaptive filtering algorithms for room acoustic signal enhancement}} \bibinfo{journal}{Signal Processing} \textbf{88}(3), \bibinfo{pages}{594--611}.

\bibitem[{Venkatakrishnan \emph{et~al.}(2013)Venkatakrishnan, Bouman, and Wohlberg}]{venkatakrishnan2013plug}
\bibinfo{author}{Venkatakrishnan, S.~V.}, \bibinfo{author}{Bouman, C.~A.},  and \bibinfo{author}{Wohlberg, B.} (\textbf{\bibinfo{year}{2013}}). \enquote{\bibinfo{title}{Plug-and-play priors for model based reconstruction}} in \emph{\bibinfo{booktitle}{2013 IEEE global conference on signal and information processing}}, \bibinfo{organization}{IEEE}, pp. \bibinfo{pages}{945--948}.

\bibitem[{Xiang(2020)}]{xiang2020model}
\bibinfo{author}{Xiang, N.} (\textbf{\bibinfo{year}{2020}}). \enquote{\bibinfo{title}{Model-based {Bayesian} analysis in acoustics---{A} tutorial}} \bibinfo{journal}{The Journal of the Acoustical Society of America} \textbf{148}(2), \bibinfo{pages}{1101--1120}.

\bibitem[{Ye \emph{et~al.}(2024)Ye, Lin, Han, Xu, Liu, Liang, Ma, Zou, and Ermon}]{ye2024tfg}
\bibinfo{author}{Ye, H.}, \bibinfo{author}{Lin, H.}, \bibinfo{author}{Han, J.}, \bibinfo{author}{Xu, M.}, \bibinfo{author}{Liu, S.}, \bibinfo{author}{Liang, Y.}, \bibinfo{author}{Ma, J.}, \bibinfo{author}{Zou, J.~Y.},  and \bibinfo{author}{Ermon, S.} (\textbf{\bibinfo{year}{2024}}). \enquote{\bibinfo{title}{{Tfg}: Unified training-free guidance for diffusion models}} \bibinfo{journal}{Advances in Neural Information Processing Systems} \textbf{37}, \bibinfo{pages}{22370--22417}.

\bibitem[{Zhang \emph{et~al.}(2021)Zhang, Li, Zuo, Zhang, Van~Gool, and Timofte}]{zhang2021plug}
\bibinfo{author}{Zhang, K.}, \bibinfo{author}{Li, Y.}, \bibinfo{author}{Zuo, W.}, \bibinfo{author}{Zhang, L.}, \bibinfo{author}{Van~Gool, L.},  and \bibinfo{author}{Timofte, R.} (\textbf{\bibinfo{year}{2021}}). \enquote{\bibinfo{title}{Plug-and-play image restoration with deep denoiser prior}} \bibinfo{journal}{IEEE Transactions on Pattern Analysis and Machine Intelligence} \textbf{44}(10), \bibinfo{pages}{6360--6376}.

\bibitem[{Zhu \emph{et~al.}(2023)Zhu, Zhang, Liang, Cao, Wen, Timofte, and Van~Gool}]{zhu2023denoising}
\bibinfo{author}{Zhu, Y.}, \bibinfo{author}{Zhang, K.}, \bibinfo{author}{Liang, J.}, \bibinfo{author}{Cao, J.}, \bibinfo{author}{Wen, B.}, \bibinfo{author}{Timofte, R.},  and \bibinfo{author}{Van~Gool, L.} (\textbf{\bibinfo{year}{2023}}). \enquote{\bibinfo{title}{Denoising diffusion models for plug-and-play image restoration}} in \emph{\bibinfo{booktitle}{Proceedings of the IEEE/CVF conference on computer vision and pattern recognition}}, pp. \bibinfo{pages}{1219--1229}.

\end{thebibliography}

\end{document}